\documentclass[12pt]{article}
\usepackage{epsfig}
\usepackage{amsfonts}
\usepackage{amsmath}

\newcommand{\be}{\begin{equation}}
\newcommand{\ee}{\end{equation}}
\newcommand{\ba}{\begin{eqnarray}}
\newcommand{\ea}{\end{eqnarray}}
\newcommand{\bi}{\begin{itemize}}  
\newcommand{\ei}{\end{itemize}}

\newcommand{\Acal}{{\mathcal A}}

\newcommand{\f}{\frac}

\newcommand{\aslash}[1]{\,\,{\raise.15ex\hbox{/}\mkern-12mu #1}}
\newcommand{\bslash}[1]{\,\,{\raise.15ex\hbox{/}\mkern-9mu #1}}

\renewcommand{\bar}{\overline}
\renewcommand{\tilde}{\widetilde}
\renewcommand{\hat}{\widehat}

\newcommand{\sect}[1]{\section{#1}\setcounter{equation}{0}}

\newcommand\lrpar{\raise .8ex\hbox{$^\leftrightarrow$} \hspace{-9pt}
\partial}
\newcommand\lpar{\raise .8ex\hbox{$^\leftarrow$} \hspace{-9pt}
\partial}
\newcommand\rpar{\raise .8ex\hbox{$^\rightarrow$} \hspace{-9pt}
\partial}

\newcommand{\gsim}{\lower.7ex\hbox{$\;\stackrel{\textstyle>}{\sim}\;$}}
\newcommand{\lsim}{\lower.7ex\hbox{$\;\stackrel{\textstyle<}{\sim}\;$}}
\begin{document}

\baselineskip=18pt

\setcounter{footnote}{0}
\setcounter{figure}{0}
\setcounter{table}{0}

\begin{titlepage}

{\begin{flushright}
 {\bf      NSF-KITP-09-110}
\end{flushright}}

\begin{center}
\vspace{1cm}

{\Large \bf  Holography from Conformal Field Theory}

\vspace{0.8cm}

{\bf Idse Heemskerk$^1$, Joao Penedones$^2$, Joseph Polchinski$^2$, James Sully$^1$}

\vspace{.5cm}

{\it $^1$ Department of Physics, University of California, \\ Santa Barbara, California
93106, USA}

{\it $^2$ Kavli Institute for Theoretical Physics \\Santa Barbara,
California 93106-4030, USA}

\end{center}
\vspace{1cm}

\begin{abstract}
The locality of bulk physics at distances below the AdS length scale is one of the remarkable aspects of AdS/CFT duality, and one of the least tested.  It requires that the AdS radius be large compared to the Planck length and the string length.  In the CFT this implies a large-$N$ expansion and a gap in the spectum of anomalous dimensions.  We conjecture that the implication also runs in the other direction, so that any CFT with a large-$N$ expansion and a large gap has a local bulk dual.  For an abstract CFT we formulate the consistency conditions, most notably crossing symmetry, and show that the conjecture is true in a broad range of CFT's, to first nontrivial order in $1/N^2$: in any CFT with a gap and a large-$N$ expansion, the four-point correlator is generated via the AdS/CFT dictionary from a local bulk interaction.  We establish this result by a counting argument on each side, and also investigate various properties of some explicit solutions.
\end{abstract}

\bigskip
\bigskip


\end{titlepage}

\sect{Introduction}

AdS/CFT duality maps a higher dimensional bulk into a lower dimensional boundary\cite{juan,GKP,Witten}. This means that excitations that are coincident in the boundary CFT may be far apart in the bulk.  From the point of view of the CFT they should be able to interact directly, but the bulk picture makes it clear that they cannot.  The purpose of this paper is to develop a better understanding of this bulk locality from the boundary CFT point of view.

In Sec.\ 2 we discuss general issues.  We first review the distinction between locality at the AdS-radius scale and at sub-AdS distances.  The former is reasonably apparent in the CFT, while the latter, which is implied by AdS/CFT duality, is remarkable and mysterious.  We then review the use of scattering experiments to probe the locality of bulk physics, and to relate it to the form of the CFT four-point function.  We examine the extent to which current tests of AdS/CFT duality probe sub-AdS locality.
Finally, we make a conjecture, to the effect that bulk locality follows from the existence of a large gap in the spectrum of operator dimensions.

In Sec.\ 3 we develop the general constraints acting on CFT's, most notably the operator product expansion (OPE) and crossing, with the goal of either showing that they imply the conjecture, or finding a counterexample.  We then specialize to a theory having only one low-dimension single-trace operator, a scalar, with a ${ \mathbb{Z}}_2$ symmetry.  This is not a full-fledged quantum field theory, in that it does not have an energy-momentum tensor, but after a thorough study of this system it will be straightforward to extend the results to scalar correlators in a complete CFT.

In Sec.\ 4 we set out to solve the constraints to first order in $1/N^2$.  An infinite set of solutions is generated via the AdS/CFT dictionary, starting from a local four-scalar interaction in the bulk.  In this context, the locality conjecture asserts that {\it all} solutions are obtained in this way.  For solutions whose intermediate spins are bounded above we show by a counting argument that our conjecture holds.   We obtain some explicit solutions in $d=2$ and $d=4$.

In Sec.\ 5 we switch to the bulk point of view.  We calculate four-point amplitudes arising from various local bulk interactions, resolve them into partial waves, and show that the results agree with those found from the abstract CFT conditions.  In Sec.\ 6 we identify the Lorentizian CFT singularity associated with bulk locality, and show that it arises from the sum over partial waves.  

In Sec.\ 7 we discuss issues of convergence of the sum over solutions.  We show that all solutions are obtained as limits of the bounded-spin solutions found earlier, and so the conjecture is true in this model.  In Sec.\ 8 we extend the model by dropping the ${ \mathbb{Z}}_2$ symmetry, and by the addition of the energy-momentum tensor to the system.  With this, the conjecture is shown to hold in a rather general set of CFT's.  We also discuss issues with higher dimension single-trace operators.

In Sec.\ 8 we discuss future directions and implications.  Our work closes off a potential loophole in the AdS/CFT correspondence, relating a mysterious property of the four-point function to an intuitive property of the spectrum of operator dimensions.  It provides a derivation of the low energy sector of AdS/CFT duality, from the assumptions of a large-$N$ expansion and gap in the spectrum of dimensions, but without an explicit string construction.  Thus it may be of use in applications such as condensed matter systems and cosmological spacetimes.

Various technical results are collected in the appendices.

\sect{General considerations}
\subsection{Coarse and sharp locality}

When the separation $l$ of the excitations in the bulk is larger than the AdS radius $R$, it is not so hard to understand why they do not interact in the CFT.  Focusing on the AdS factor, it is the radial direction $r$ that is emergent.  There is an approximate identification of gauge theory energy with radius,
\begin{equation}
E \sim r/R^2\ . \label{Erholo}
\end{equation}
This arises from the warping of the space: the same bulk excitation, at different $r$, will have different energies as seen in the gauge theory.  Locality in $r$ then follows from locality in CFT energy, as exhibited for example by the renormalization group.

This effect is actually a bit of a red herring, as we will soon explain, but it is interesting to explore it further.  Does locality in energy mean, for example, that a high-energy particle will pass through one's body?  Remarkably, yes, if it is prepared correctly.  (We thank Lenny Susskind for asking this, and for volunteering to test the effect.)  A proton from a high energy beam or a cosmic source would scatter inelastically and shower, and increasing its energy would just lead to a bigger shower.  However, if the proton is created by a local operator not too far away from Lenny, and with sufficiently large boost, then its constituents will spread very little before reaching him.  From the point of view of low energy fields, it is then nearly invisible: since it is a color singlet, its leading nuclear interaction is its parametrically small dipole moment, and so for large enough boost it will pass through him without interaction.  This is known as color transparency~\cite{{Bertsch:1981py}}.  We should also worry about the electromagnetic interaction, but we can suppress this in the same way by considering a charge zero state, for example an atom, again producing all constituents at a point.   In term of AdS/CFT duality, what is happening is that the particle is produced at the $r = \infty$ boundary, and passes by Lenny before falling into the IR where he lives.\footnote{The connection between color transparency and holography was pointed out by Matt Strassler (private communication).}  Notice that what plays the role of $E$ in Eq.~(\ref{Erholo}) is really the inverse size of the system, its internal state, and not its center of mass energy.

We can expect this locality in energy only to hold approximately, to $\delta E/E = O(1)$, and so locality in $r$ holds only to a resolution $\delta r/r = O(1)$.  Given the radial part of the AdS metric, $ds^2 = R^2 dr^2/r^2$, this implies a spatial resolution down to $l \sim R$.  However, AdS/CFT implies much more.  We expect local field theory in the bulk to hold at least down to the string scale $l_{\rm s}$, while the AdS scale $R = \lambda^{1/4} l_{\rm s}$ is parametrically larger in the regime where there is a bulk spacetime of small curvature.

In discussing scalings we are focusing on the $AdS_5 \times S^5$ case, but the principle is general.  We should note that the existence of the bulk spacetime requires both large $g^2 N = \lambda $ and large  $ g^{-2} N = N^2/\lambda $  (by S-duality), so $N$ must also be large.    In other words, the string scale can never be larger than the Planck scale.  It can be equal, or absent as in M theory duals, in which case the Planck scale $l_{\rm P}$ governs both the large-$N$ expansion and the gap in dimensions.  In particular, the AdS radius is $l_{\rm P} $ times a power of $N$. 

Thus, energy-radius holography nicely explains part of the emergence of the bulk spacetime, but also misses a critical aspect.  The existence of locality down to a fixed physical scale that can be parametrically smaller than the AdS length remains a mystery in the CFT.  Thus, we refer to {\it coarse holography} and {\it sharp holography}, and it is the latter that we seek to explain.\footnote{In an earlier version we referred to these as `horizon' and `sub-horizon' locality.  We thank John McGreevy for pointing out that this is poor terminology.}

It has been argued that sharp holography emerges from the matrix structure of the gauge theory, e.g.~\cite{Susskind:1998vk,Asplund:2008xd}.  The Eguchi-Kawai model~\cite{Eguchi:1982nm}  may be a useful parallel, in that spacetime emerges from the color structure on a single site.  In the present work, however, this color structure will not play a direct role.  We will be focusing on the large anomalous dimensions at strong coupling, and will not inquire into their origin in the gauge theory dynamics.  It would be valuable to have a clearer understanding of their dynamical origin, and this may well involve the color structure.

\subsection{Scattering and the four-point function}
\label{AdSscat}
The locality properties can be translated into a quantitative statement about gauge theory amplitudes via a scattering thought experiment~\cite{Polchinski:1999ry,Susskind:1998vk,FSS,Gary:2009ae,Gary:2}.\footnote{For other approaches to this question see Refs.~\cite{holo,small}.}
Consider the CFT four-point function 
\begin{equation}
\left\langle \prod_{i=1}^4 {\cal O}(t_i,\hat{e}_i) \right\rangle\ ,
\end{equation}
where the CFT lives on [Lorentzian time $\times$ $S^3$].  From the bulk point of view, the CFT operators create and destroy excitations at the boundary.  Taking $t_{1,2}$ to the past of $t_{3,4}$, we can think of  this as corresponding to a 2-to-2 scattering process.  In order to probe the locality structure, we must convolve with sources $f_i$ so as to focus the excitations into narrow beams,
\begin{equation}
{\cal A} = \left\langle \prod_{i=1}^4 {\int dt_i\, d^3\hat e_i\, f_i(t_i, \hat e_i)\cal O}(t_i,\hat{e}_i) \right\rangle\ .
\end{equation}
For large $R$ the packets can have both small $\delta t_i$ and $\delta e_i$ and also an energy-momentum spread that is small compared to the mean value in the packet.  They then focus in a region small compared to the AdS length.

\begin{figure} 
 \centering
\includegraphics[width=7cm]{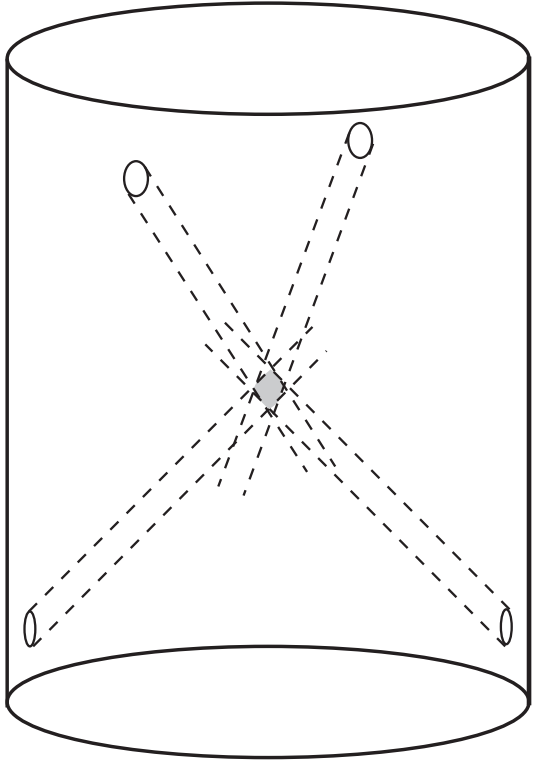}
\caption{Four-point correlator with wavepackets aligned to intersect in the bulk.}
\label{scatter}
\end{figure}
  Now, if the beams intersect, as in Fig.~\ref{scatter}, the amplitude ${\cal A}$ will be large.  However, if we change slightly the location or momentum of a source, then the beams will miss, and the amplitude will drop off rapidly.  From the point of view of the bulk theory this is clear, but again in the boundary it is mysterious: it appears that a small change in the form of ${\cal A}$ leads to a large change in its value.\footnote{Refs.~\cite{FSS,Gary:2} identify `backgrounds' to the thought experiment, and refine the conditions that must be satisfied by the wavepackets, but they do not contradict the assertion that one can probe physics below the AdS scale.  In particular, the backgrounds discussed in Ref.~\cite{Gary:2} drop out in the flat spacetime limit defined in Ref.~\cite{Polchinski:1999ry,Susskind:1998vk}.}

This property can be nicely recast as the statement that the CFT four-point function has a certain singularity when the operators are aligned so as to allow a classical bulk scattering process to occur~\cite{Gary:2009ae}.  For example, if the initial particles start diametrically opposite and with equal energies at time $t_{1,2} = -\pi/2$, they will meet in the center of global AdS at $t=0$, scatter into new (but still opposite) directions, and reach diametrically opposite points on the boundary at time $t_{3,4} = +\pi/2$.
The four-point function is therefore singular when the arguments are $(-\pi/2, \hat{e})$, $(-\pi/2, -\hat{e})$, $(+\pi/2, \hat{e}')$, $(+\pi/2, -\hat{e}')$, and in all conformally equivalent configurations.

This singularity is not present in general CFT's, for example not in the weakly coupled ${\cal N}=4$ theory (there is a weaker singularity at the same point).  
Rather, it emerges in the strong-coupling limit.  In Sec.~6 we will describe the singularity in more detail, and compare it with what we find in the CFT.  For now, the main lesson is that to study the bulk locality properties we should look at the CFT four-point function.  Note that the forms of the two- and three-point functions are fully determined by conformal symmetry, but that of the four-point function is not.  In fact, in all dimensions it is determined by symmetry up to a function of two real cross ratios.  This function carries dynamical information, in particular regarding the locality of the bulk theory.

\subsection{Current understanding}

AdS/CFT duality has been subjected to many tests.  Indeed, every time we apply it in a new way we have the possibility that it will lead to implausible or incorrect results, signaling a failure of the duality.  The tests are of many types, for example
\begin{itemize}
\item
BPS states and amplitudes
\item
Behavior under symmetry breaking and RG flow
\item
Calculations using integrability
\item
Numerical tests, both by light-cone methods and by Monte Carlo
\item
Comparison with experiment, in heavy ion physics
\end{itemize}
This list is not exhaustive, and in each category there are many separate tests.  So we can ask, do any of these test sharp holography?  Studies of BPS states and amplitudes, integrability calculations, and numerical tests are thus far limited to the spectrum of operator dimensions, and the two- and three-point functions, and do not probe the four-point function where locality becomes visible.  Renormalization group behavior depends only on the energy-radius relation associated with AdS scale holography, and applications to real QCD would seem to be too coarse to distinguish sub-AdS scales.

Thus, a possible way in which AdS/CFT duality might break down is through the failure of sub-AdS holography~\cite{FSS}.\footnote{We thank M. Douglas and S. Giddings for discussions of this possibility.}  Perhaps the symmetries of theory, together with energy-radius holography, imply only that the gauge theory reconstructs some version of the bulk string theory that is smeared over the AdS radius.  Thus, by investigating the constraints imposed by the axioms of conformal field theory, we expect either to identify what such a smeared theory might be, or to exclude this possibility.

We should note some other possible tests of sub-AdS locality that have been suggested to us.  The recent application of AdS/CFT duality to the ${\cal N}=4$ S-matrix~\cite{Alday:2007hr} involves detailed properties of higher-point functions.\footnote{We thank S. Dubovsky and D. Gross for this observation.}    However, the only amplitudes for which there is a fairly strong all-orders understanding are the on-shell four and five point functions \cite{Bern:2005iz}, which are highly constrained by symmetry \cite{Drummond:2007cf}, whereas we need at least the off-shell four-point function.  The scattering of brane probes, rather than supergravity excitations, might make locality more manifest.\footnote{We thank E. Silverstein for raising this point.}    For example, in the BFSS matrix model such scattering is used even in the weakly coupled limit to study the gravitational force law.  However, we note that even in this model nontrivial strong coupling effects are needed to ensure the duality, and so we suspect that there must be some issue analogous to what we study.  Finally, if the famous factor of 3/4 in the partition function~\cite{Gubser:1996de} were confirmed by gauge theory reasoning, it would imply that the bulk theory is sensitive to the precise form of the Schwarzschild metric, even below the AdS scale.\footnote{We thank L. Susskind for this observation.}

\subsection{A conjecture}

The AdS/CFT dictionary~\cite{juan,GKP,Witten} relates the dimension of any CFT operator to the mass of the corresponding bulk state,
\begin{equation}
\Delta(\Delta-4) = m^2 R^2\ .
\end{equation}
Kaluza-Klein excitations have masses of order $1/R$, and so their dimensions are of order one.  String excitations have masses of order $1/l_{\rm s} = \lambda^{1/4}/R$, and so their dimensions are of order $\lambda^{1/4}$.  To have local physics below the AdS scale, we must have a hierarchy between the AdS and string scales, or $\lambda \gg 1$.  In this case, the great majority of operators, all of those dual to string excitations, have parametrically large dimensions.  

This gap in dimensions is a striking and unusual property of the CFT.  We see that it is implied by sharp holography, and it is natural to conjecture that the implication runs in the other direction as well.  That is, in any CFT in which most operators get parametrically large anomalous dimensions, there will be a bulk dual with sub-AdS locality.  A plausible precise statement is this: {\it Conjecture: any CFT that has a large-$N$ expansion, and in which all single-trace operators of spin greater than two have parametrically large dimensions, has a local bulk dual.}

The discussion in Sec.~2.1 in the $d=4$, ${\cal N}=4$ context shows that sub-AdS locality is possible only when $N$ is parametrically large.  More generally, bulk locality implies that the AdS radius must be large compared to the Planck length, and so there must be some expansion parameter in the CFT that corresponds to the gravitational loop expansion in the bulk.  We will refer to this as a large-$N$ expansion, even in the absence of an explicit Lagrangian description of the \mbox{CFT}.  
In particular, the bulk loop expansion allows us to distinguish single-particle states from multi-particle states, and so we designate the corresponding operators as single-trace and multi-trace by analogy with the known examples.\footnote{A large-$N$ expansion usually implies that the Hamiltonian is itself a single-trace operator, and we make this assumption explicit in order to exclude known exceptions in which a CFT is coupled to itself or to an otherwise-decoupled CFT through multi-trace interactions~\cite{Aharony:2001pa}.  This assumption will make an interesting appearance in Sec.~4.3.3.
We thank E. Silverstein and O. Aharony for emphasizing these exceptions.}  The essential property of the expansion is that the connected expectation values of single-trace operators are suppressed compared to the disconnected expectation values.\footnote{Note that this is true for large-$N$ vector models as well as matrix models, but that vector models do not have large gaps in their dimension.  Thus the latter are expected to have duals in which quantum gravitational effects are small but string effects are of order one~\cite{Klebanov:2002ja}.}

The single-trace condition picks out operators that are dual to single-particle states.  
It would seem to be necessary that all single-trace operators of spin greater than two have large dimension, because we know of no low-energy effective theory that would be a candidate to describe their bulk physics.\footnote{In AdS there are consistent interacting theories of higher spin fields (see \cite{Bekaert:2005vh} for a recent review). However, these theories are non-local because they contain an infinite tower of fields with increasing spin and interactions with unbounded number of derivatives.}  So we are conjecturing that the strongest necessary condition that we can identify is actually the sufficient condition.  If this is true, then the condition for sub-AdS locality is reduced from a rather mysterious property of the four-point function to a much more intuitive property of the operator dimensions, which can be determined from the two-point function.

\sect{CFT constraints}

\subsection{Generalities}

Conformal field theories are constrained by the operator product expansion (OPE), conformal invariance, crossing, unitarity, and modular invariance.\footnote{For classic work on this subject see Refs.~\cite{classic}.  For applications to AdS/CFT correlators see Refs.~\cite{AdSOPE}.}
The general form of the OPE is
\begin{equation}
{\cal O}_i(x) {\cal O}_j(0) = \sum_{k} x^{\Delta_k - \Delta_i - \Delta_j}  c^{k}\!_{ij} {\cal O}_k(0)
\end{equation}
(Lorentz indices suppressed).  This has a finite radius of convergence in any correlator, given by the distance to the nearest other operator.  The OPE coefficients $c^{k}\!_{ij}$ and the operator dimensions $\Delta_i$ can thus be regarded as the data defining the CFT, subject to the other constraints. 

We are interested in CFT with a parameter $\lambda$, such that some operator dimensions become parametrically large while the remainder have a finite limit.  Interior to the radius of convergence $x_c$, the total contribution of the high-dimension operators is parametrically suppressed as $(x/x_c)^{\Delta_{\rm large}}$.  We will thus study the limiting theory, in which the low-dimension operators have a closed operator algebra among themselves.

Using the OPE twice reduces the four-point function to the two-point function:\begin{equation}
{\cal A}_{ijkl} = \sum_{m} c^{m}\!_{ij }c_{mkl} =  \sum_{m} c^{m}\!_{ik} c_{mjl}\ . \label{ope}
\end{equation}
Indices are lowered with the two-point function, which can be given a conventional normalization; we have suppressed the coordinate dependence for simplicity.   The OPE can be applied in two ways with overlapping radii of convergence, as indicated.  The crossing condition is the equality of these two sums, and it is a strong constraint on the $c^{i}\!_{jk}$.

Unitarity would give positivity conditions on the terms in the sums~(\ref{ope}).  However, this will not be of use to us, because we will be working in the $1/N^2$ expansion.  The leading terms will be manifestly positive, but the first correction, where the issue of bulk locality arises, can have either sign.  

Modular invariance will also not be useful to us, for two reasons.  First, it relates low-dimension operators to high-dimension operators, as in the Cardy relations, while we are interested in the present work only in relations among the low-dimension operators.  Second, the set of local operators in a 4d CFT is isomorphic to the gauge theory quantized on $S^3$.  The partition function would then be given by the amplitude on $S^3 \times S^1$.  Here there is no large diffeomorphism interchanging the spatial and time directions, to give a modular invariance relation.  We could consider instead the CFT on $T^3 \times S^1$, for example, where there would be constraints from large diffeomorphisms, but then there is no direct connection to the spectrum of operators.

In summary, we will be solving the constraints arising from the OPE, conformal invariance, and crossing on the algebra of low dimension operators.

\subsection{A scalar model}

The OPE governs the behavior of the four-point correlator when any two operators become coincident.  Our assumption about operator dimensions translates directly into information about these limits.  Bulk locality, on the other hand, is related to a singularity in this correlator at Lorentzian points corresponding to classical scattering.  These points are not conformally equivalent to those governed by the OPE, as we will review in Sec.~6, so there is not an immediate link between our assumption and the result that we hope to derive from it.  Rather, we must use the other constraints in combination with our assumption in order to derive general restrictions on the correlator.  As we will explain, this will in effect require that we find the general solution to the crossing condition.

The idea of finding all CFT's subject to general consistency conditions is an old one.  It has been realized most fully in the case of rational conformal field theory, where there is the additional assumption of an extended conformal symmetry such that the spectrum contains only a finite number of irreducible representations.  In our case, the additional assumption is a restricted set of low-dimension operators.

In the simplest CFT, the only low dimension single-trace operator would be the energy-momentum tensor, so that the corresponding bulk dual would involve only gravity.\footnote{Note that our focus is orthogonal to that in Ref.~\cite{Witten:2007kt}.  That paper is largely concerned with high-dimension black hole states, which we have decoupled, while the 2+1 dimensional bulk has no light propagating fields.  Correspondingly all correlators of the energy-momentum tensor in that work are immediately determined by holomorphy.  However, there may be an interesting story that includes both directions.}  However, we will take an even simpler model, in which the only low dimension single-trace operator is a scalar ${\cal O}$ of dimension $\Delta$.   After a thorough study of the crossing constraint in this system, it will be quite simple to include also the energy-momentum tensor in the OPE, and so constrain the scalar correlator in a full-fledged CFT.

As an aside, a CFT without an energy-momentum tensor would seem to be an oxymoron.  What it is missing is an operator that could evolve the CFT state from one time to the next; it is a set of correlators without a notion of causality.  We must measure the boundary state at {\it every} time in order to reconstruct the bulk state at a {\it single} time, and so there is no holography, as should be expected for a theory without gravity in the bulk.  This does not affect its use as warmup for us, as the form of the crossing condition is very similar to that in a full CFT.  This model could actually arise as a sector of an AdS compactification in which there is a light scalar with self-interaction much stronger than gravity, working in the approximation that gravity decouples.

We will further assume a ${ \mathbb{Z}}_2$ symmetry ${\cal O} \to - {\cal O}$, so that the operator ${\cal O}$ does not itself appear in the ${\cal O}{\cal O}$ OPE.  The lowest dimension operator in the OPE, aside from the unit operator, is then the double trace ${\cal O}^2$, with dimension $2\Delta + O(1/N^2)$.  All other double-trace operators are obtained by differentiating one or the other of the ${\cal O}$ in ${\cal O}^2$.
Total derivatives generate conformal descendant operators, whose contribution is determined by symmetry in terms of those of the primary operators.  To list all primary operators we need consider only the difference $\lrpar = \rpar-\lpar$ acting between the two ${\cal O}$'s.  A complete set of primary double-trace operators is
\begin{equation}
{\cal O}_{n,l} \equiv 
{\cal O} \lrpar_{\mu_1} \ldots  \lrpar_{\mu_l}  (  \lrpar_{\nu} \,  \lrpar^{\nu })^n {\cal O} - {\rm traces} \ ,
\label{double}
\end{equation}
such as to be traceless on the $\mu$'s.   This has spin $l$ and dimension $\Delta_{n,l} = 2\Delta + 2n + l + O(1/N^2)$.

The contribution of higher-trace operators in the OPE is absent at the order in $1/N^2$ in which we work.  We normalize ${\cal O}$ to be $1/N$ times a trace of adjoint variables, so that the two-point function and disconnected four-point function are of order $N^0$.  The connected four-point function is of order $1/N^2$.\footnote{The operator ${\cal O}$ can get an anomalous dimension at order $1/N^2$. This will give a correction to the disconnected four-point function that is of the same order as the leading connected contribution. However, crossing does not mix these two $1/N^2$ contributions to the four-point function and so we ignore the disconnected correction.}
The double-trace ${\cal O}^2$ appears in the $\cal O \cal O$ OPE at order $N^0$, and the square of this term gives the disconnected four-point function.  The leading connected contribution comes from the order $1/N^2$ correction to this OPE coefficient, times the $N^0$ coefficient.  It also gets a contribution from the order $1/N^2$ shift of the dimensions of the double-trace operators.   Triple-trace terms do not enter due to the ${ \mathbb{Z}}_2$ symmetry, and quadruple-trace terms enter first at order $1/N^4$.

\subsection{Constraints in the scalar model}

We consider the Euclidean four-point function.  Using conformal invariance, we can bring the operators to lie in a single plane, with complex coordinate $z$, and then to the standard positions 0, 1, $\infty$, $z$.  This is familiar from the string world-sheet, but in any dimension there are two independent cross-ratios, which can be combined into $z$.

The remaining information from conformal invariance determines the contributions of the descendants in terms of the primaries, and so the four-point function is represented in terms of a sum over primaries:
\ba
\langle {\cal O}(0)  {\cal O}(z,\bar z) {\cal O}{(1)} {\cal O}(\infty) \rangle
&\equiv& {\cal A}(z,\bar z) 
=  \frac{1}{(z\bar z)^\Delta}
+\sum_{n=0}^\infty \sum_{l=0}^\infty p(n,l) \, \frac{g_{\Delta(n,l),l} (z,\bar{z})}{(z\bar z)^\Delta}\ .
\label{CPW}
\ea
(In taking $z_4 \to \infty$, a factor of $(z_4\bar z_4)^\Delta$ is implicitly introduced to give a smooth limit.)   The first term corresponds to the unit operator (vacuum), and the sum runs over the primary double trace operators~(\ref{double}).  Here $p(n,l)$ is the square of the OPE coefficient, and $g_{\Delta(n,l),l} (z,\bar{z})$ are the conformal blocks, representing the total contribution of the conformal family over the given primary.

For $d=2$~\cite{Dolan:2000ut},
\ba
g_{E,l}(z,\bar{z}) =\f{ k(E+l,z)\, k(E-l,\bar{z})+k(E+l,\bar{z})\, k(E-l,z)}{1+\delta_{l,0}} \label{CPW2d}
\ea
with
\ba
k(\beta,z)=z^\f{\beta}{2} F_{\beta/2}(z)\ ,\quad  F_{\beta/2}(z)=F\left(\f{\beta}{2},\f{\beta}{2},\beta,z\right)\ ,
\ea
where $F(a,b,c,z)$ denotes the standard hypergeometric function $\,_2F_1$.
For $d=4$~\cite{Dolan:2000ut},
\ba
g_{E,l}(z,\bar{z}) =\f{z\bar{z}}{z-\bar{z}}\left[ k(E+l,z)\, k(E-l-2,\bar{z})-k(E+l,\bar{z})\, k(E-l-2,z)\right]\ .\label{CPW4d}
\ea
Note that in the $d=2$ case only the $SO(2,2)$ symmetry is being used and not the full Virasoro algebra, which would be much more constraining.  Consequently, the crossing condition is very similar in $d=2$ and $d=4$, and we will study the $d=2$ case first because it is slightly simpler.

The crossing condition implies that $l$ must be even (from interchanging the vertex operators at $z$ and $0$ or at 1 and $\infty$ via a conformal transformation), and also that
\begin{equation}
 {\cal A}(z,\bar z) =  {\cal A}(1-z,1-\bar z)\ . \label{crossing}
\end{equation}
This is the only remaining condition to be satisfied.

As we have discussed, we will necessarily be solving in the $1/N$ expansion, 
\ba
{\cal A}(z,\bar{z})  &=& \Acal_0(z,\bar{z})  + \frac{1}{N^2}\Acal_1(z,\bar{z})  + \ldots\ , \nonumber\\
p(n,l) &=& p_0(n,l) + \frac{1}{N^2} p_1(n,l) + \ldots\ , \nonumber\\
\Delta(n,l) &=& 2\Delta + 2n + l +  \frac{1}{N^2} \gamma_1(n,l) + \ldots \ .
\ea
Then
\be
(z\bar z)^\Delta {\cal A}_0(z,\bar z)  =  1
+\sum_{n=0}^\infty \sum_{l=0 \atop \rm even}^\infty p_0(n,l) \,{g_{2\Delta + 2n + l,l} (z,\bar{z})} \ ,
\ee
and
\be
(z\bar z)^\Delta {\cal A}_1(z,\bar z)  = 
\sum_{n=0}^\infty \sum_{l=0 \atop \rm even}^\infty p_1(n,l) \,{g_{2\Delta + 2n + l,l} (z,\bar{z})}
+  p_0(n,l) \gamma_1(n,l) \, \frac{1}{2}\frac{\partial}{\partial n} {g_{2\Delta + 2n + l,l} (z,\bar{z})}\ .
\label{A1cross}
\ee

\sect{Solving the constraints}

\subsection{General considerations}

How constraining is crossing?\footnote{We should note the recent papers \cite{Rattazzi:2008pe,Rychkov:2009ij}, which derive useful constraints on operator dimensions from general CFT properties.  These papers use the constraint from unitarity, so we will not be able to apply the same approach, but they provide a nice example of the power of the crossing condition.}
  The unknowns $p(n,l)$ and $\Delta(n,l)$ are indexed by two integers.  Equation (\ref{crossing}) is a function of one complex or two real variables.  Equivalently, by analytic continuation in the real and imaginary parts of $z$ we can regard $z$ and $\bar z$ as independent, and the crossing equation is a holomorphic function of two variables.  By expanding in suitable complete sets of functions of $z$ and of $\bar z$ we get equations indexed by two integers.  Thus there seems to be a rough equality between the number of equations and the number of unknowns, but whether the number of solutions is zero, finite, or infinite depends on the detailed structure.

In the $1/N$ expansion, in fact, we can readily identify an infinite number of solutions.  If we have a scalar field theory in the bulk of AdS space, it defines boundary correlators via the AdS/CFT dictionary~\cite{juan,GKP,Witten}.  These are conformally invariant and Bose symmetric, and by inserting a complete set of intermediate states in the bulk we can derive the \mbox{OPE}.  Therefore these correlators at tree level will satisfy all of our conditions to order $1/N^2$.  A $\lambda \phi^4$ bulk interaction, or any quartic interaction with additional derivatives, will provide a solution (a trilinear interaction is forbidden by the ${ \mathbb{Z}}_2$ symmetry).  Going to higher orders in $1/N^2$ we will encounter nonrenormalizable divergences in the bulk, since we have decoupled the stringy physics, but at each order we can define a renormalized solution at the cost of new parameters, representing the unknown contribution of the high-dimension operators.

Any such solution will be local by construction.  In Sec.~6 we will verify that these satisfy the formal locality condition of \cite{Gary:2009ae}.  Further, it is expected that {\it any} local theory can be derived from a local Lagrangian, at least if it has a weak coupling expansion as is necessarily the case here.  It follows that one form of our locality conjecture is simply that the solutions constructed from local bulk Lagrangians make up {\it all} solutions of the crossing condition.  In the current setting our goal is to show this, or to find a counterexample.

We would like to be able to just count solutions.  For example, we could restrict to bulk interactions with a maximum number of derivatives, and make the corresponding restriction on the CFT, and then see if the solutions are equal in number.  The independent interactions with up to six derivatives are $\phi^4$, $\phi^2 \phi_{;\mu\nu} \phi^{;\mu\nu}$, and  $\phi^2 \phi_{;\mu\nu\sigma}
 \phi^{;\mu\nu\sigma}$, using integration by parts and the equations of motion.  The AdS curvature tensor is expressed in terms of the metric, so we need not include interactions involving the background curvature.  As we will show in Sec.~5, an interaction with $2k$ derivatives in the bulk corresponds to perturbations $p_1(n,l)$, $\gamma_1(n,l)$ growing as $n^{2k + {\rm const.}}$ in the CFT.  Thus we could count the number of solutions whose large-$n$ behavior is bounded by a given power, and compare with the number of interactions with the corresponding number of derivatives.   
 
 In fact it will be simpler to look at the spins of the intermediate states.  The interaction $\phi^4$ destroys and creates only two-particle states of spin 0, and so the corresponding $p_1(n,l)$, $\gamma_1(n,l)$ are nonzero only for $l = 0$.  The interaction  $\phi^2 \phi_{;\mu\nu} \phi^{;\mu\nu}$ creates and destroys two-particle states of spin 0 or 2, as does $\phi^2 \phi_{;\mu\nu\sigma} \phi^{;\mu\nu\sigma}$ (note that the spin must be even, by Bose symmetry).  Thus, we will count bulk interactions according to the maximum spin that they can couple to, and similarly will count CFT solutions by the maximum value of $l$ for which the perturbation is nonzero.  In Sec.~7 we will discuss the completeness of the solutions found in this way.
 
Let us now count the bulk interactions with four scalars plus derivatives.  Counting such operators up to total derivatives and equations of motion is equivalent to counting flat space $S$-matrices.  We can count these by starting with monomials in the Mandelstam variables, $s^a t^b u^c$ with $a \geq b \geq c$, and Bose-symmetrizing.  However, operators proportional to $s + t + u = 4m^2$ (where $m^2$ is an arbitrary scalar mass) are not independent.   Given any operator with highest monomial $s^a t^b u^c$, we can obtain a dependent operator containing the monomial $s^{a+1} t^b u^c$ by multiplying by $s + t + u$.  The only operators that cannot be obtained in this way are those whose highest monomial has $a = b$.  Thus, a complete set is obtained from the $a+1$ monomials $s^a t^a u^c$ such that $a \geq c \geq 0$.  Further, the total spin is maximized in the $u$-channel, where each $s$ or $t$ can give rise to a factor of $\cos \theta$.  Thus the operator couples to maximum spin $2a$.  In all, there are $a + 1$ interactions of maximum spin $2a$. These have $2k$ derivatives, for $k = 2a, 2a+1, \ldots , 3a$.  The total number of interactions with spin at most $L$ is then $
\sum_{a = 0}^{L/2} (a+1) = (L+2)(L+4)/8$.

\begin{figure} 
 \centering
\includegraphics[width=12cm]{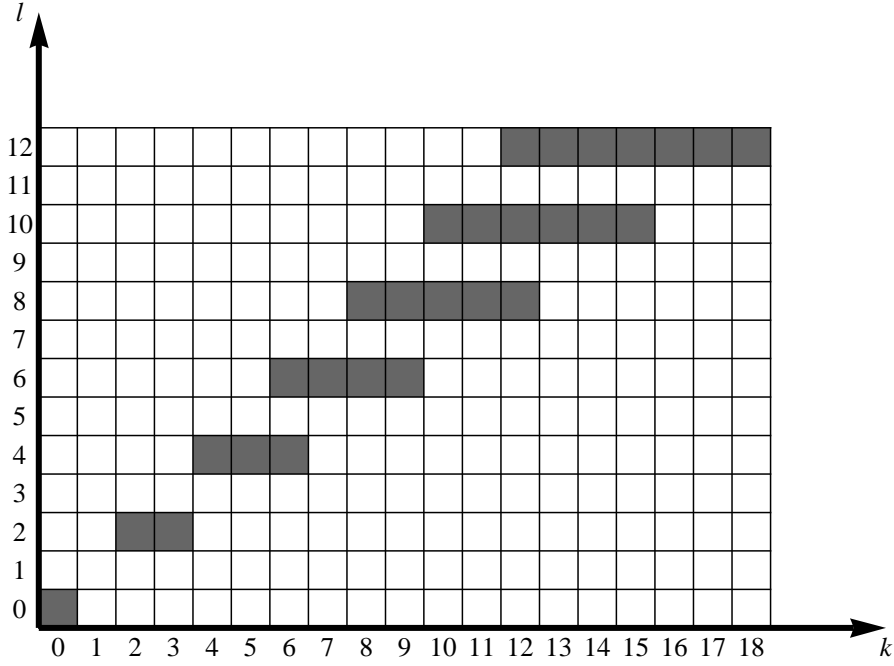}
\caption{ Quartic interactions of spin $l=2a$ and with $2k$ derivatives. There are $1+a$ interactions of even spin $l=2a$, with the number of derivatives given by $k = 2a, 2a+1, \ldots , 3a$.
The total number of interactions with spin at most $L$ is $(L+2)(L+4)/8$.}
\label{tabela}
\end{figure}

These facts are summarized in Fig.~\ref{tabela}.  The interactions discussed above are built from the monomials 1, $st$, and $stu$ and correspond to the 3 gray squares closer to the bottom left corner of Fig.~\ref{tabela}. 

\subsection{Counting solutions in $d=2$}

The $N^0$ correlator is
\ba \label{freecorrelator}
\Acal_0(z,\bar{z})=  \frac{1}{(z\bar{z})^\Delta} + \frac{1}{[(1-z)(1-\bar{z})]^\Delta}  + 1\ .
\ea
By Taylor expanding the conformal partial wave expansion~(\ref{CPW}) around $z,\bar{z}=0$ one finds that 
\ba
p_0(n,l)=\left[1+(-1)^l\right] C_n C_{n+l}\ ,\quad C_n =  \f{\Gamma^2(\Delta+n)\Gamma(2\Delta+n-1)}{n! \Gamma^2(\Delta)\Gamma(2\Delta+2n-1)} \label{3pt2D}
\ .
\ea

It is convenient to rewrite the $O(1/N^2)$ condition~(\ref{A1cross}) as
\be
 {\cal A}_1(z,\bar z)  = 
\sum_{n=0}^\infty \sum_{l=0 \atop \rm even}^\infty \left[ c(n,l) \, h_{n,l}(z,\bar{z}) + \gamma (n,l)\, \frac{\partial}{2\partial n} h_{n,l}(z,\bar{z}) \right] \label{CPWexp}\ ,
\ee
where
\ba
h_{n,l}(z,\bar{z}) &=& \frac{ p_0(n,l)}{1+\delta_{l,0}}  (z \bar z)^n\left(
z^{l} F_{\Delta + n + l}(z) F_{\Delta + n}(\bar z) +  \bar z^{l} F_{\Delta + n}(z) F_{\Delta + n+l}(\bar z)  \right)\ ,\nonumber\\
c(n,l) &=& \frac{2 p_1(n,l) - \gamma_1(n,l) \partial_n p_0(n,l)
}{2 p_0(n,l)}
\ ,\quad  \gamma(n,l) = \gamma_1(n,l)\ .
\ea

There is a branch cut $\ln(z\bar z)$ at the point $z, \bar z=0$ from $ \partial_n$ acting on $ (z \bar z)^n$.  
At $z,\bar z =1$ (which get mapped to the origin on the other side of the crossing relation), there are also logarithms, from the hypergeometric functions. 
One way to organize the conditions is to look near $z = 0$ and $\bar z = 1$ (remember, we can analytically continue independently in both variables), and at each point separate things into one term which is a log times a holomorphic function, and a second which is holomorphic.  The crossing relation ${\cal A}(z,\bar z) =  {\cal A}(1-z,1-\bar z)$ thus becomes four holomorphic equations, which appear multiplying
$\ln z   \ln(1-\bar z) $, $\ln z$, $\ln(1-\bar z)$, and 1.

We must be careful because the radius of convergence of the sums is 1 in $z, \bar z$, from the coincidence of operators, and so additional singularities could arise there from the sums.  By taking $z$ to be small, we regulate both the sums over $n$ and $l$ for the first term in $h_{n,l}$, but only the sum over $z$ in the second term.  To proceed at this point we introduce at this point the restriction explained in the previous subsection, that $l$ is bounded above by some given $L$, in order to exclude additional $\ln(1-\bar z)$ behavior.  In Sec.~7 we will see that this is not actually necessary. 

It will be sufficient to ignore terms holomorphic in $1- \bar z$, keeping only the holomorphic terms that multiply $\ln(1-\bar z)$ in the crossing relation:
\begin{eqnarray}
\sum_{n=0}^\infty \sum_{l=0 \atop \rm even}^L \left[ c(n,l)  + \gamma (n,l)\, \frac{\partial}{2\partial n}  \right] 
\left\{ \frac{ p_0(n,l)}{1+\delta_{l,0}} \left[ z^{n+l} \bar z^{n} F_{\Delta + n + l}(z) \tilde F_{\Delta + n}(1-\bar z) + ( n \leftrightarrow n+ l) \right]  \right\}&&
\nonumber\\
\qquad\qquad = \frac{1}{2}
\sum_{n=0}^\infty \sum_{l=0 \atop \rm even}^L \gamma (n,l) h_{n,l} (1-z,1-\bar{z})  \ .
\qquad\qquad 
&& \label{ln1zb}
\end{eqnarray}
On the left-hand side we have used
\begin{eqnarray}
F_{\beta/2}(\bar z) &=& \ln(1-\bar z)\tilde F_{\beta/2}(1- \bar z) + {{\rm holomorphic\ at}\ \bar z = 1}\ ,
\nonumber\\
\tilde F_{\beta/2}(z)  &=& - \frac{\Gamma(\beta)}{\Gamma^2(\beta/2)} F(\beta/2,\beta/2,1,z)\ .
\end{eqnarray}
Similarly, the holomorphic terms that multiplies $\ln z$ in Eq.~(\ref{ln1zb}) give
\begin{eqnarray}
&& \sum_{n=0}^\infty \sum_{l=0 \atop \rm even}^L \gamma (n,l)\, 
\frac{ p_0(n,l)}{1+\delta_{l,0}} \left[ z^{n+l} \bar z^{n} F_{\Delta + n + l}(z) \tilde F_{\Delta + n}(1-\bar z) + ( n \leftrightarrow n+ l) \right] 
\label{lnln}\\
 &=& \sum_{n=0}^\infty \sum_{l=0 \atop \rm even}^L \gamma (n,l)\, 
 \frac{ p_0(n,l)}{1+\delta_{l,0}} \left[ (1-z)^{n+l} (1-\bar z)^{n} \tilde F_{\Delta + n + l}(z) F_{\Delta + n}(1-\bar z) +  ( n \leftrightarrow n+ l) \right] \ .
 \nonumber 
\end{eqnarray}

To simplify further we can project onto a complete set $z^{\Delta + n} F_{\Delta + n}(z)$.  These are eigenfunctions,
\begin{equation}
\label{eigen}
{\cal D}\, z^{\Delta + n} F_{\Delta + n}(z) = (\Delta + n)(\Delta + n -1) z^{\Delta + n} F_{\Delta + n}(z)\ ,\quad
{\cal D} = z^2 \partial_z (1-z) \partial_z\ .
\end{equation}
It follows that
\begin{equation}
\oint_C \frac{dz}{2\pi i} z^{m - m' - 1}  F_{\Delta + m}(z)  F_{1 -\Delta -m'}(z) = \delta_{m,m'}\ , \label{proj}
\end{equation}
where the contour $C$ circles 0 counterclockwise, and the normalization is readily obtained by a Laurent expansion of the integrand.  We define
\begin{equation}
J(m,m') = \f{C_{m}}{C_m'} I(m,m')\ ,\quad I(m,m') = 
\oint_C \frac{dz}{2\pi i} \frac{(1-z)^{m}}{z^{m'+1}}  \tilde F_{\Delta + m}(z)  F_{1 -\Delta -m'}(z) \ . \label{defJ}
\end{equation}
Projecting thus onto $m' = p$ around $z=0$ and onto $m' = q$ around $\bar z =1$, the crossing condition~(\ref{lnln}) becomes
\begin{equation}
\sum_{l=0 \atop \rm even}^L  \gamma(p,l) J(p+l,q) + \sum_{l=2 \atop \rm even}^L  \gamma(p-l,l) J(p-l,q)
= ( p \leftrightarrow q)\ . \label{pq}
\end{equation}
Notice that this only involves the anomalous dimension $\gamma(n,l)$ and not the coefficient $c(n,l)$.  
This condition is symmetric in $p,q$ and trivial for $p=q$.

As discussed in Sec.~4.1, we want to count solutions where the sum over $l$ is limited to a finite range $\leq L$.  As a warmup, consider the case of $L=0$, where the relations~(\ref{pq}) become
\be
\gamma(p,0) J(p,q) = \gamma(q,0) J(q,p)\ . \label{pql0}
\ee
Letting $q = 0$, we immediately obtain the solution
 \be
\gamma(p,0) = \gamma(0,0) \frac{J(0,p)}{J(p,0)}=\gamma(0,0)\frac{2\Delta-1}{2\Delta+2p-1}\ ,\label{L=0}
\ee
where $J(p,0)$ and $J(0,p)$ are obtained in appendix \ref{J}.
Thus there is at most one solution for $\gamma(n,0)$ when $L=0$, determined up to the overall normalization $\gamma(0,0)$.  Indeed, it seems remarkable that there are any solutions at all, since we must satisfy Eq.~(\ref{pql0}) for all $q$.  The consistency of the solution requires that
\begin{equation} 
\frac{J(p,q)}{J(q,p)} 
=\frac{2\Delta+2p-1}{2\Delta+2q-1} \label{consist}
\end{equation}
for all $p$ and $q$.  This relation is not at all obvious from the definition.  Nevertheless, we know that there must be at least one solution, corresponding to the bulk interaction $\phi^4$, and indeed the consistency condition (\ref{consist}) holds in the cases we have checked.  Having determined $\gamma(n,0)$, we can immediately determine $c(n,0)$, because this appears only on the left-hand side of Eq.~(\ref{ln1zb}), and the $c(n,0)$ for different $n$ multiply independent functions.  Thus there is exactly one solution with $L=0$.

This example illustrates the strategy we will use for general $L$.  The crossing relation~(\ref{pq}) appears to overdetermine $\gamma(n,l)$, and gives an upper bound on the number of solutions; for each solution of~(\ref{pq}), $c(n,l)$ is uniquely determined by equation (\ref{ln1zb}).  The counting of bulk interactions gives a lower bound, and we will show that these bounds are equal, thus determining the actual number of solutions.

\begin{figure}
 \centering
\includegraphics{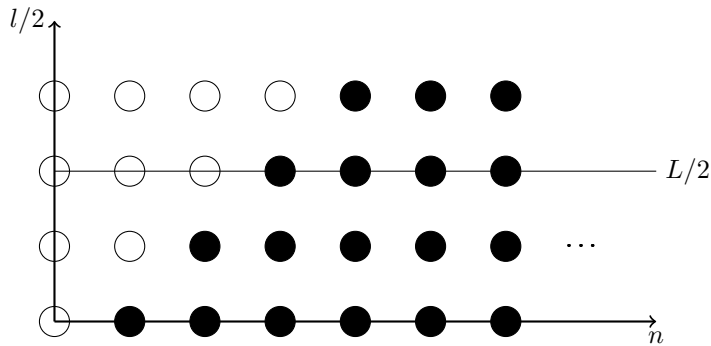}
\caption{Open circles are $\gamma(n,l)$ not determined by the equations while filled circles correspond to $\gamma(n,l)$ determined by the equations (\ref{pq}) with $p=0,1,\dots,n$ and $q=0,1, \dots,{\rm min}(p-1,L/2)$. We see that up to spin $L$ we have $(L/2+1)(L/2+2)/2$ undetermined $\gamma(n,l)$.}
\label{countfig}
\end{figure}

We can use the $(p,q)=(1,0)$ condition to solve for $\gamma(1,0)$ in terms of all the other $\gamma(0,l), \gamma(1,l)$.  We can then use $(p,q)=(2,0), (2,1)$ to solve for $\gamma(2,0), \gamma(2,2)$ in terms of all the other $\gamma$ with $p \leq 2$.  Proceeding in this way, at fixed $p$ we use the conditions at $0 \leq q \leq{\rm{min}}(p-1,L/2)$ to solve for $\gamma(p,0), \ldots, \gamma(p,{\rm{min}}(2p-2,L))$ in terms of the remaining $\gamma(p',q)$ at $p' \leq p$.    In order for the solution at each step to exist and be unique, we need that $M(p)_{qr} = J(p + 2r, q)$  for $r,q = 0, 1, \ldots, k-1$ be a nondegenerate $k \times k$ matrix.  This is shown in Appendix~\ref{mqr}.  In the end we have solved for all $\gamma(n,l)$ with $l < 2n$, so the solution is determined by specifying the $\gamma(n,l)$ for $2n \leq l \leq L$, as depicted in Fig. \ref{countfig}.  The total number of free parameters is $\sum_{p = 0}^{L/2} (\frac{1}{2} L + 1 - p) = (L+2)(L+4)/8$.  
For each solution, the $c(n,l)$ for different $n,l$ multiply independent functions of $z,\bar z$ in Eq.~(\ref{ln1zb}), and so are uniquely determined.  Thus there are at most $(L+2)(L+4)/8$ solutions to the crossing condition with maximum spin $L$.

This is the same as our count of interactions in Sec.~4.1: our upper and lower bounds agree.  We can conclude that the total number of solutions of the crossing relations is exactly equal to the number of bulk interactions, and our conjecture is true for interactions restricted to bounded $L$.  In Sec.~7 we will argue that all solutions can be obtained as convergent sums of such bounded solutions.  We can readily extend the analysis of crossing to include the energy-momentum tensor in the \mbox{OPE}.  However, in this case the solutions necessarily involve all $l$, and so we defer this until Sec.~8.

\subsection{More on the $d=2$ solutions }

\subsubsection{Large-$q$ equations}

The CFT axioms allow us not only to count the solutions, but to construct them, as we have done above for $L=0$.  Here we will extract a few results from the crossing equation~(\ref{pq}) in the limit  $q \gg p,\Delta, L$. Using Eqs.~(\ref{plargeq},  \ref{qlargeq}), the leading terms on the two sides of the crossing equation give
\be
\frac{ (2\Delta + 2p+2L-1)B_{p+L}}{B_p}{\gamma(p,L) }
\approx q^{1-2L} \left( \sum_{l=0 \atop \rm even}^L \gamma(q,l) 
+ \sum_{l=2 \atop \rm even}^L \gamma(q-l,l) \right) \ , \label{leadLHS}
\ee
where $B_p$ is defined in Eq.~(\ref{plargeq}).
We have divided through by $-B_p q^{2\Delta + 2p + 2L -1}$, with the result that the LHS depends only on $p$ and the RHS only on $q$.  On each side the corrections are of relative order $1/q$ and generically depend on $p,l,L$.  If $\gamma(p,L)$ is nonzero, it follows that at least for some $l\le L$,  $\gamma(q,l)$ must grow at large $q$ as $q^{2L-1}$ or faster.  If the leading terms do not cancel, so the RHS has a nonzero large-$q$ limit $K$, then $\gamma(q,l)$ actually saturates this lower bound on its asymptotic behavior and  we obtain an explicit solution
\begin{equation}
 \gamma(p,L)= K \frac{B_p}{(2\Delta+2p+2L-1) B_{p+L} }\ .\label{gammaL}
\end{equation}
It is easy to check that  (\ref{gammaL}) indeed behaves as  $p^{2L-1}$ for large $p$.
Furthermore, setting $L=0$ we recover the unique solution (\ref{L=0}).
Eq.~(\ref{gammaL}) only gives the anomalous dimensions for the maximal spin. 
The $\gamma(p,l)$ for $l<L$ can be obtained from the subleading terms in the large $q$ expansion of  (\ref{pq}), though this is cumbersome.

As we shall see in section \ref{localizing}, a bulk interaction with $2k$ derivatives gives rise to anomalous dimensions $\gamma(p,l)$ going like $p^{2k-1}$  at large $p$.
Therefore, the result  (\ref{gammaL})  corresponds to bulk interactions of spin $L$ and with $2L$ derivatives, i.e. the leftmost box in each row in Fig. \ref{tabela}. The other boxes in each row require cancelations among the leading terms of the large $q$ expansion of the RHS of (\ref{pq}). 
Consider the second box in each row, i.e.  an interaction with spin $L$ and $2L+2$ derivatives. Then, the leading behavior of LHS of (\ref{pq}) is (\ref{leadLHS}), as before, but the RHS naively grows as $q^{2\Delta+2p+2L}$.
Vanishing of the terms of order $q^{2\Delta+2p+2L}$ and $q^{2\Delta+2p+2L-1}$ requires
\begin{equation}
\lim_{q \to \infty}\sum_{l=0 \atop \rm even}^L \f{ \gamma(q,l) }{(1 + \delta_{l,0})q^{2L+1}}=0\ ,\label{cond1}
\end{equation}
and
\begin{equation} 
\lim_{q \to \infty}\sum_{l=0 \atop \rm even}^L \f{ \gamma(q,l) \, l}{(1 + \delta_{l,0})q^{2L+1}}=
\frac{2}{2L+1}\lim_{q \to \infty}\sum_{l=0 \atop \rm even}^\infty \f{ \gamma(q,l) }{(1 + \delta_{l,0})q^{2L}} \ .\label{cond2}
\end{equation}
The terms of order $q^{2\Delta+2p+2L-2}$ then provide an explicit solution for the anomalous dimensions
\begin{equation}
 \gamma(p,L)= K' \frac{\left[L(2L+1)+ (2\Delta+2p-1) (2\Delta+2p+2L-1) \right]B_p}{(2\Delta+2p+2L-1) B_{p+L} }\ ,\label{gammaLsub}
\end{equation}
where
\begin{equation}
 K'=2 \lim_{q \to \infty}\sum_{l=0 \atop \rm even}^L \f{\gamma(q,l) \,l^2}{(1 + \delta_{l,0})q^{2L+1}} \ .
\end{equation}
Notice that this solution only exists for $L\ge 2$ in agreement with Fig. \ref{tabela}. Indeed, conditions (\ref{cond1}) and (\ref{cond2}) for $L=0$ imply that $\gamma(q,0)$ behaves as $q^{-1}$ for large $q$ and therefore the bulk interaction can not contain any derivatives.

\subsubsection{The coefficients $c(n,l)$}

We shall now consider the computation of the coefficients $c(n,l)$. 
These are uniquely fixed by equation (\ref{ln1zb}) given the anomalous dimensions $\gamma(n,l)$.
Let us illustrate this statement with the $L=0$ example. In this case, it is enough to keep the terms of  order $(1-\bar{z})^0$ in (\ref{ln1zb}),
\be
-\sum_{n=0}^\infty \left[ 2 c(n,0) +\gamma(n,0) \frac{\partial}{\partial n}\right] C_n^2 \frac{\Gamma(2\Delta+2n)}{\Gamma^2(\Delta+n)}z^n F_{\Delta+n}(z)
=\gamma(0,0)  F_{\Delta}(1-z)\ .
\ee
It is convenient to rewrite this equation in the following form
\be
-2\sum_{n=0}^\infty c'(n,0)  C_n^2 \frac{\Gamma(2\Delta+2n)}{\Gamma^2(\Delta+n)}z^n F_{\Delta+n}(z) = H(z)\ , \label{cprime}
\ee
where\footnote{The solutions $\gamma(n,l)$ are always rational functions of $n$, and this defines the extension to non-integer $n$ as required to take the $n$-derivative.}
\be
c'(n,0)=c(n,0)- \frac{1}2{\partial_n}\gamma(n,0) \ ,
\ee
and
\be
H(z)= \gamma(0,0)  F_{\Delta}(1-z)+
\sum_{n=0}^\infty \frac{\partial}{\partial n}\left(\gamma(n,0)  C_n^2 \frac{\Gamma(2\Delta+2n)}{\Gamma^2(\Delta+n)}z^n F_{\Delta+n}(z)\right)\ ,
\label{Hofz}
\ee
which is entirely determined given the anomalous dimensions $\gamma(n,0)$. 
Applying the projector (\ref{proj}) to equation (\ref{cprime}) one determines the coefficients $c'(n,0)$.
Surprisingly, these coefficients are all zero. In fact, using the explicit solution (\ref{L=0}) for the anomalous dimensions one finds that the function $H(z)$ vanishes identically (see appendix \ref{Hz}).
Thus, we conclude that 
\ba
c(n,l)=\frac{1}{2}\f{\partial}{\partial n} \gamma(n,l)\ . \label{cdg}
\ea
Notably, this statement holds true in all the examples we have considered, though we have not been able to find a general derivation and had to rely on "empirical" data.
It has been previously observed~\cite{Cornalba:2006xm} that the unitarity bound on the leading term of the OPE in the crossed channel implies (\ref{cdg}) for large $n,l$. 
However, it was also shown in Ref.~\cite{Cornalba:2006xm}  that the conformal partial wave expansion of a scalar exchange in AdS only respected (\ref{cdg}) asymptotically. 
We suspect that  (\ref{cdg}) only holds for amplitudes corresponding to bulk contact interactions.

\subsubsection{Multi-trace interactions}

At special values of $\Delta$, multi-trace interactions become marginal.  We have excluded these in our conjecture, but they should still appear as solutions to the crossing relation.  The first examples in $d=2$ are ${\cal O}^2$ at $\Delta = 1$, ${\cal O}^3$ at $\Delta = 2/3$,  ${\cal O}^4$ at $\Delta = 1/2$, and ${\cal O} \lrpar_{\nu} \,  \lrpar^{\nu } {\cal O}$ at $\Delta = 0$.  The bilinear interactions ${\cal O}^2$ and ${\cal O} \lrpar_{\nu} \,  \lrpar^{\nu } {\cal O}$ just shift $\Delta$ in the disconnected amplitude, which trivially respects crossing.  The interaction ${\cal O}^3$ is forbidden by ${ \mathbb{Z}}_2$, so the first interesting case is ${\cal O}^4$ at $\Delta = 1/2$.

This operator couples only to spin 0, for which we have the solution~(\ref{L=0}),
\be
\gamma(p,0) = \frac{C}{2\Delta+2p-1}\ , \quad c(p,0) = -\frac{C}{(2\Delta+2p-1)^2}\ .
\ee
As $\Delta \to 1/2$ at fixed $C$, $\gamma(0,0)$ and $c(0,0)$ diverge.  The double poles cancel in the expression for the correlator, but a single pole remains.  There are two ways that we might think to obtain a finite limit.  We could let $C$ scale as $2\Delta - 1$, so that the only nonzero term is $n=l=0$; we will call this the first solution.  To obtain the second solution, we could hold $C$ finite but subtract off the pole, which is proportional to the first solution.  However, this second solution is not conformal: the limiting process introduces logs of the separations, just as in the usual dimensional regularization.

This has a simple interpretation.  When a multi-trace interaction becomes marginal, the single-trace interaction sources the multi-trace interaction under RG flow~\cite{Tseytlin:1999ii}.  We identify this flow as the second solution, while the first solution corresponds to keeping only the quadruple-trace interaction and tuning the single-trace interaction to zero.  We can confirm this by calculating directly the effect of the marginal perturbation $\mathcal{O}^4$ in the CFT:
\be
\mathcal{A}_1 \propto \int \frac{d^2 w}{|w||1-w||z-w|}
 = \pi^2 \left[ F_{\frac{1}{2}}(z) F_{\frac{1}{2}}(1-\bar{z})+F_{\frac{1}{2}}(1-z)F_{\frac{1}{2}}(\bar{z})\right]\ . \label{marg}
\ee
This suggests that the conformal partial wave decomposition of the connected four-point function 
only contains the $l=n=0$ term.
Indeed, when  $\Delta=1/2$, we see that expression (\ref{L=0}) gives $\gamma(n,0)=0$ for all $n>0$ and
 the four point function (\ref{marg}) can be written using a single partial wave,
\be
\int \frac{d^2 w}{|w||1-w||z-w|} = 
 -\frac{\pi}{2} \left.\frac{\partial}{\partial n}  h_{n,0}(z,\bar{z})\right|_{n=0 \ \atop \Delta \to \frac{1}{2}} \ .
\ee
The marginal behavior of ${\cal O}^4$ is lifted both by $1/N^2$ effects and by second order effects in the ${\cal O}^4$ interaction, giving a coupled RG flow for the two interactions~\cite{Tseytlin:1999ii}.

\subsection{Counting Solutions in $d=4$}

In $d=4$ the different form of the conformal blocks (\ref{CPW4d}) leads to slightly more complicated results, but the procedure for finding solutions is analogous.
As in $d=2$, we can find the $N^0$ coefficients in the partial wave expansion from matching to the $N^0$ correlator (\ref{freecorrelator}). They are
\begin{eqnarray} \label{p0d4}
 p_{0}(n,l) &&=\left[1+(-1)^l\right]\frac{2(l+1)(2\Delta+2n+l-2)}{(\Delta-1)^2} C^{(\Delta-1)}_{n}C^{(\Delta-1)}_{n+l+1}\ ,
\end{eqnarray}
where $C^{(\Delta-1)}_{n}$ is the coefficient from the two-dimensional case with $\Delta$ replaced by $\Delta-1$.
Keeping only the terms multiplying $\log z\log(1-\bar z)$ in the equation $\mathcal{A}_1(z,\bar z) = \mathcal{A}_1(1-z,1-\bar z)$, where $\mathcal{A}_1(z,\bar z)$ is written as in (\ref{CPWexp}), we find
\begin{eqnarray}
&& \sum_{n=0}^\infty \sum_{l=0 \atop \rm even}^L \gamma (n,l)\, 
p_0(n,l)\frac{\bar z}{1-\bar z}\left[ z^{n+l} \bar z^{n-1} F_{\Delta + n + l}(z) \tilde F_{\Delta + n-1}(1-\bar z) - ( n-1 \leftrightarrow n+ l) \right] \label{lnln4}
\\
 &=& \sum_{n=0}^\infty \sum_{l=0 \atop \rm even}^L \gamma (n,l)\, 
p_0(n,l)\frac{z-1}{z}\left[ (1-z)^{n+l} (1-\bar z)^{n-1} \tilde F_{\Delta + n + l}(z) F_{\Delta + n-1}(1-\bar z) - ( n-1 \leftrightarrow n+ l) \right] 
 \nonumber
\end{eqnarray}
Again we project twice using (\ref{proj}): onto $m'=p-1$ around $z=0$ and onto $m'=q-1$ near $\bar z =1$. We define $J^{(\Delta-1)}(p,q)$ as $J(p,q)$ with the shift $\Delta \rightarrow \Delta-1$.
We also define 
\begin{equation}
\gamma'(n,l) = \frac{2(l+1)(2\Delta+2n+l-2)}{(\Delta-1)^2} \gamma(n,l)
\, .
\end{equation}
In terms of these functions, the crossing condition becomes
\begin{equation} \label{crossingeqns4d}
\sum_{l=0 \atop \rm even}^L
\Big[\gamma'(p-l-1,l)J^{(\Delta-1)}(p-l-1,q)-\gamma'(p,l)J^{(\Delta-1)}(p+l+1,q)\Big] = (p\leftrightarrow q)
\, .
\end{equation}

Exactly analogously to the two-dimensional case, we consider the crossing relation at fixed $p$ for a series of $q$ such that $0 \leq q \leq \min(p-1,L/2)$. We use these relations to solve for $\gamma(p,0), \, ...\, , \, \gamma(p,\min(2p-2,L))$ in terms of all of the remaining $\gamma(p',l)$ for $p'<p$. In the four-dimensional case, the existence and uniqueness of solutions requires that $M^{(\Delta-1)}(p)_{qr} = J^{(\Delta-1)}(p+2r+1,q)$ for $r,q=0,1,\,...\,,\,k-1$ is a non-degenerate $k\times k$ matrix. This is shown in appendix \ref{nondegen}. As the counting is the same as in two-dimensions, we again have that there are at most $(L+2)(L+4)/8$ solutions to the crossing relation with maximum spin $L$. 

\subsection{More on $d=4$ Solutions}

Two differences complicate the four-dimensional crossing relation as compared to the two-dimensional case. First, the two terms on either side have different arguments even when $l=0$. Consequently, it is not possible to obtain a solution for $\gamma(p,0)$ in a simple way, even though $J^{(\Delta-1)}(p,q)$ has the symmetry (\ref{consist}). Second, the relative minus sign between the two terms on either side of the equation causes the highest powers in $p$ to cancel. We are then forced to look at subleading terms to find a solution for $l=L$.

The first complication can be overcome by brute force. The $L=0$ crossing condition is
\begin{equation} \label{crossingL0d4}
\gamma'(p-1,0)J^{(\Delta-1)}(p-1,q)-\gamma'(p,0)J^{(\Delta-1)}(p+1,q) = ( p \leftrightarrow q ) 
\end{equation}
Setting $q=0$ the overlap integrals can actually be computed (see appendix \ref{J}) and lead to a recursion relation for the $\gamma(p,0)$. The relation is solved by\footnote{
As in $d=2$, the case $\Delta=1$ in $d=4$ is special because $\mathcal{O}^4$ can generate a marginal deformation of the CFT. }
\begin{equation} \label{gammaL0}
\gamma(p,0)= \frac{(2\Delta+p-3)(p+1)(\Delta+p-1)(2\Delta-1)}{(\Delta-1)(2\Delta+2p-3)(2\Delta+2p-1)}\gamma(0,0)  \ .
\end{equation}

That leaves the second complication. For two dimensions a solution for $l=L$ was found by equating the coefficients of the highest power of $p$ (or $q$) on both sides of the crossing condition. As in two-dimensions, using \eqref{plargeq} we can find the large $p$ limit of the overlap integral:  
\begin{equation} \label{d4largep}
 J^{(\Delta-1)}(p,q) = -2B^{(\Delta-1)}_q p^{2\Delta+2q-1} + O((1/p)^{-2\Delta-2q+2})\, ..
\end{equation}
However, the sign difference between the two terms in the crossing condition makes the leading $p$ terms on the left hand side of (\ref{crossingeqns4d}) cancel and we have to look at subleading terms. There are subleading terms of two types; denoting the relative order in $1/p$ by a parenthesised superscript, 
\begin{equation}
 J^{(\Delta-1)}(p,q) = \left[J^{(\Delta-1)}(p,q)\right]^{(0)} + \left[J^{(\Delta-1)}(p,q)\right]^{(1)} + ...   \,\, ,
\end{equation}
these are
\begin{align}
 \left[J^{(\Delta-1)}(p+x,q)\right]^{(0)}&=\left[J^{(\Delta-1)}(p,q)\right]^{(0)}\ , \\
 \left[J^{(\Delta-1)}(p+x,q)\right]^{(1)} &= \partial_p \left[J^{(\Delta-1)}(p,q)\right]^{(0)} x  + \left[J^{(\Delta-1)}(p,q)\right]^{(1)}\ .
\end{align}
Only the subleading terms that come as derivatives of the leading terms are relevant. The other subleading terms cancel out of the crossing equation and thus we will not need to expand $J$ to higher order in $1/p$ to find them.

We can now use (\ref{plargeq}) on the LHS and (\ref{qlargeq}) on the RHS to determine the leading $p$ term of (\ref{crossingeqns4d}). On the RHS we get \begin{equation}
\gamma'(q,L)B^{(\Delta-1)}_{q+L+1}(2\Delta+2q+2L-1)p^{2q+2L+2\Delta-2} 
\end{equation}
while on the LHS
\begin{eqnarray}
2 B^{(\Delta-1)}_q p^{2\Delta+2q-4} \sum_{l=0 \atop \rm even}^L (l+1)\Big(2\gamma'^{(0)}(p,l)(2\Delta+2q-3)+p \partial_p\gamma'^{(0)}(p,l)\Big) \ .
\end{eqnarray}
We see that the lower bound on large $p$ growth of $\gamma(p,l)$ is now $p^{2L+1}$ ($\gamma'(p,l)$ is now $p^{2L+2}$), in accordance with what we find in section \ref{localizing}.
Assuming the absence of cancelations among the leading terms so that the bound is saturated we have $\gamma^{'(0)}(q,l)=\alpha(l)q^{2L+2}$ so
\begin{equation}
4 B^{(\Delta-1)}_q p^{2\Delta+2q-4} (2\Delta+2q+L-2) \sum_{l=0 \atop \rm even}^L (l+1)\alpha(l)=\gamma'(q,L)B^{(\Delta-1)}_{q+L+1}(2\Delta+2q+2L-1)p^{2q+2L+2\Delta-2} 
\end{equation}
from which we find the solution for maximal spin
\begin{equation} \label{d4maximalspingamma}
\gamma(m,L) = \frac{K}{2\Delta+2m+2L-1}\frac{B^{(\Delta-1)}_{m}}{B^{(\Delta-1)}_{m+L+1}}\ , \quad  K = \frac{2}{L+1}\sum_{l=0 \atop \rm even}^L (l+1)\alpha(l)\ .
\end{equation}

Again as in two dimensions, we find an equation analagous to (\ref{Hofz}) that determines $c(n,l)$ in terms of a finite subset of $\gamma(m,l)$. Using a computer algebra program, we are able to calculate a finite number of $c(n,0)$ and find in all cases that
\be
c(n,l) = \frac{1}{2}\frac{\partial}{\partial n} \gamma(n,l)
\, .
\ee


\sect{Bulk calculations}

\subsection{Small-$L$ examples}

\begin{figure} 
 \centering
\includegraphics[width=5cm]{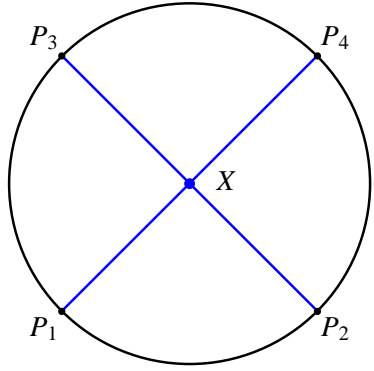}
\caption{ Witten diagram for the CFT four-point function associated to a quartic contact interaction in AdS.
The boundary points $P_i$ are connected to the bulk interaction point $X$ by bulk to boundary propagators.
In general, the quartic vertex at $X$ includes derivatives acting on the bulk to boundary propagators.}
\label{AdScontact}
\end{figure}

We have seen that the number of solutions to the CFT constraints precisely matches the counting of quartic local bulk interactions.
We now wish to illustrate this equivalence more explicitly. We shall start by computing some simple Witten diagrams, as in Fig. \ref{AdScontact}, describing four-point functions of the dual CFT. Four point functions obtained in this way will automatically satisfy all the CFT constraints. 
The goal of this section is to relate these correlators with the solutions found in the previous section.

Let us consider Euclidean AdS$_{d+1}$ defined by the hyperboloid
\be
X^2=-R^2=-1\ ,\ \ \ \ \ \ \ \ \ \  X^0 > 0\ ,\ \ \ \ \ \ \ \ \ \ \ X \in \mathbb{M}^{d+2}\ ,
\ee
embedded in $(d+2)$-dimensional Minkowski spacetime.
We shall  set $R=1$ in this and the following section.
It is convenient to think of the conformal boundary of AdS as the space of null rays 
\footnote{See \cite{Penedones:2007ns} for a brief review of this formalism first proposed by Dirac \cite{Dirac:1936fq}.}
\be
P^2=0\ ,\ \ \ \ \ \ \ \ \ \  P\sim \lambda P\ \ (\lambda \in \mathbb{R})\ ,\ \ \ \ \ \ \ \ \ \ \ P \in \mathbb{M}^{d+2}\ . 
\ee
Then, the correlations functions of the dual CFT are encoded into $SO(1,d+1)$ invariant functions of the external points $P_i$, transforming
homogeneously with weights $\Delta_i$.
In particular, the general form of a four-point function of dimension $\Delta$ scalar operators is
\be
A(P_1,P_2,P_3,P_4)=\frac{\mathcal{A}(z,\bar{z})}{P_{12}^\Delta P_{34}^\Delta}\ ,
\ee
where $P_{ij}=-2P_i\cdot P_j$ is positive  for future directed $P$'s and $\mathcal{A}$ only depends on the cross ratios,
\be
 u=\frac{P_{12}P_{34}}{P_{13}P_{24}}=\frac{1}{z\bar{z}}\ ,\ \ \ \ \ \ 
v= \frac{P_{14}P_{23}}{P_{13}P_{24}}=\frac{(1-z)(1-\bar{z})}{z\bar{z}}\ .
\ee

The basic ingredient required to compute Witten diagrams is the bulk to boundary propagator which in this notation is simply given by
\be
(-2 P \cdot X)^{-\Delta}\ ,
\ee
up to a normalization constant that will not be important for us.
We are now ready to compute the four point function associated with a quartic $\phi^4$ interaction in AdS,
\be
(z\bar{z})^\Delta \Acal_1(z,\bar{z}) \propto 
P_{13}^\Delta P_{24}^\Delta \int_{\rm AdS} dX \prod_{i=1}^4 (-2 P_i \cdot X)^{-\Delta} \ .
\ee
This is precisely the definition of the reduced D-function.\footnote{See \cite{D'Hoker:1999pj, Dolan:2000ut} for D-function properties and definitions. D-functions in the embedding space notation are reviewed in appendix B of \cite{Gary:2009ae}.} 
We can then write
\be
(z\bar{z})^\Delta \Acal_1(z,\bar{z}) \propto  \bar{D}_{\Delta\Delta\Delta\Delta}(u,v) \ .
\ee
The conformal partial wave expansion of this correlator can be found using a series representation of the D-function \cite{Dolan:2000ut,Dolan:2000uw}.
In both $d=2$ and $d=4$ we recover the unique solution with $L=0$ found in the previous section, Eqs.~(\ref{L=0}) and (\ref{gammaL0}) respectively.

A quartic interaction with only 2 derivatives does not generate a new four-point function.
The vertex $\phi^2 (\nabla \phi)^2 $ can be reduced to $\phi^4$ by integrating by parts and using the equations of motion.
The first new contribution comes from an interaction vertex with 4 derivatives,
\ba
(\nabla \phi)^2 (\nabla \phi)^2 \ .
\ea
To compute the contribution of this vertex to the four-point function it is useful to introduce embedding space derivatives
\ba
\nabla_N=\partial_N+X_N\, X\cdot \partial\ .
\ea
This combination removes the radial derivative in the embedding space. More precisely, 
\ba
\nabla_N f(X^2)=2 X_N (1+X^2) f'(X^2)=0\ ,
\ea
as it should for a constant function in AdS.
The  four-point function is then given by
\ba
&&\sum_{{\rm perm\ } P_{i}}\int_{\rm AdS} dX\, \nabla_N (-2P_1\cdot X)^{-\Delta}\nabla^N (-2P_2\cdot X)^{-\Delta}
\nabla_M (-2P_3\cdot X)^{-\Delta}\nabla^M (-2P_4\cdot X)^{-\Delta} \nonumber \\
&\propto&  \int_{\rm AdS} dX\, \prod_{i=1}^4 (-2P_i\cdot X)^{-\Delta} 
\sum_{{\rm perm\ } P_{i}} \left( \frac{P_1\cdot P_2\, P_3\cdot P_4}{P_1\cdot X \,P_2\cdot X \,P_3\cdot X \,P_4\cdot X } +\dots
\right)\ ,
\ea
where the dots give rise to the same four-point function as a  $\phi^4$ interaction.
The new contribution to the reduced amplitude is
\ba
(z\bar{z})^\Delta \Acal_1(z,\bar{z})\propto (1+u+v) \bar{D}_{\Delta+1\,\Delta+1\,\Delta+1\,\Delta+1}(u,v)\ . \label{AL2k4}
\ea
The explicit conformal partial wave expansion (\ref{CPWexp}) of this correlator in $d=2$ is given in appendix \ref{expCPW}.
The expansion only contains partial waves with spin 0 and 2.
The  property (\ref{cdg}) is obeyed and the maximal spin $L=2$ anomalous dimensions are given by 
expression (\ref{gammaL}). In $d=4$ the  property (\ref{cdg}) is also verified and 
the anomalous dimensions of the maximal spin $L=2$ are given by (\ref{d4maximalspingamma}).

The next new interaction comes from an interaction vertex with 6 derivatives,
\ba
(\nabla \phi)^2 (\nabla_{\mu} \nabla_\nu \phi)^2 \ .
\ea
The four-point function is then given by
\ba
&&\sum_{{\rm perm\ } P_{i}}\int_{\rm AdS} dX\,  \nabla_N (-2P_1\cdot X)^{-\Delta} \nabla^N (-2P_2\cdot X)^{-\Delta}
\nabla_M \nabla_K  (-2P_3\cdot X)^{-\Delta}\nabla^M  \nabla^K(-2P_4\cdot X)^{-\Delta} \nonumber\\
&\propto& \int_{\rm AdS} dX\, \prod_{i=1}^4(-2P_1\cdot X)^{-\Delta}
\sum_{{\rm perm\ } P_{i}} \left( \frac{P_1\cdot P_2\, (P_3\cdot P_4)^2}{P_1\cdot X \,P_2\cdot X (P_3\cdot X )^2(P_4\cdot X)^2 }
+ \dots \right) \ . 
\ea
The dots correspond to terms that give the same as $\phi^4$ or $(\nabla \phi)^2(\nabla \phi)^2$ interactions.
The first term gives something new. Its contribution to the reduced amplitude is
\ba
(z\bar{z})^\Delta  \Acal_1(z,\bar{z})&\propto&  
\bar{D}_{\Delta+2\,\Delta+1\,\Delta+2\,\Delta+1}(u,v) + \bar{D}_{\Delta+1\,\Delta+2\,\Delta+1\,\Delta+2}(u,v) \nonumber \\
&+& 
u^2 \bar{D}_{\Delta+2\,\Delta+2\,\Delta+1\,\Delta+1}(u,v)  
 +u \bar{D}_{\Delta+1\,\Delta+1\,\Delta+2\,\Delta+2}(u,v) \label{AL2k6}\\
&+&
v^2\bar{D}_{\Delta+1\,\Delta+2\,\Delta+2\,\Delta+1}(u,v) +v \bar{D}_{\Delta+2\,\Delta+1\,\Delta+1\,\Delta+2}(u,v)\ . \nonumber 
\ea
We give the conformal partial wave expansion of this correlator in $d=2$  in  appendix \ref{expCPW}.
Again, the expansion only contains partial waves with spin 0 and 2 and (\ref{cdg}) holds.
The spin 2 anomalous dimensions in $d=2$ are now given by a linear combination of (\ref{gammaLsub}) and (\ref{gammaL}) with $L=2$.

With this examples we have explored the left bottom corner of Fig.~\ref{tabela}, which summarizes the possible bulk interactions.
We have found that solutions to the CFT constraints are in one-to-one correspondence to local bulk interactions, 
in agreement with our conjecture.
However, the complexity of the explicit computations increases very rapidly as the 
spin and the number of derivatives of the interaction grow. 
In the next subsection we shall consider an approximation scheme that allows us to determine part of the conformal partial wave
expansion for a family of bulk interactions.

%
%
%

\subsection{Regge limit}

The papers \cite{Cornalba:2006xk,Cornalba:2006xm,Cornalba:2007zb,Cornalba:2007fs,Cornalba:2008qf}
explored the regime of high energy scattering in AdS and its consequences for the CFT four-point function.
They show that this kinematical regime is sufficient to determine
the highest spin part of the conformal partial wave decomposition of the CFT four-point function.
This limit is the AdS/CFT analogue of the well known Regge limit in flat space scattering.  It will allow us to find some solutions for general $L$.

\begin{figure} 
 \centering
\includegraphics[width=4cm]{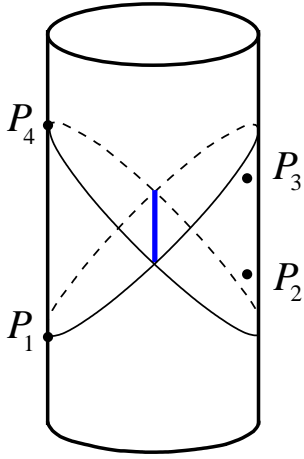}
\caption{ External points $P_i$ in the boundary of conformally compactified AdS.
The Regge limit corresponds to $P_3 \to -P_1$ or $P_2 \to -P_4$. 
In this limit the dominant contribution to the four-point function comes from the AdS region around the 
future lightcone of $P_1$ and past lightcone of $P_4$, in particular, from their intersection at the $(d-1)$-dimensional hyperboloid shown in blue.}
\label{ReggeAdS}
\end{figure}

As in flat space, the Regge limit is intrinsically Lorentzian and we must consider physical 
AdS$_{d+1} \subset \mathbb{R}^{2,d} = \mathbb{M}^2 \times \mathbb{M}^d$.
A particularly convenient way of taking the Regge limit is to choose the following external points 
\ba
&P_1=(1,0,0)\ ,\ \ \ \ \ \ \ \ \ \ \ 
&P_2=(\bar{x}^2,1,\bar{x})\ ,\\
&P_3=(-1,-x^2,x)\ ,\ \ \ \ \ \ \ \ \ 
&P_4=(0,-1,0)\ ,
\ea
where $P=(P^+,P^-,P^a) \in \mathbb{M}^2 \times \mathbb{M}^d$ with metric $dP^2=-dP^+dP^-+dP^adP_a$.
The conformal invariant cross ratios are then given by
\ba
z\bar{z}=x^2\bar{x}^2\ ,\ \ \ \ \ \ \ \ \ 
z+\bar{z}=-2 x \cdot \bar{x} \ ,
\ea
and the Regge limit corresponds to $z,\bar{z} \to 0$ with fixed ratio $\bar{z}/z$.
As depicted in Fig.~\ref{ReggeAdS}, this corresponds to the limit of small scattering angle in the bulk.
This configuration of the cross ratios can be obtained as an analytic continuation of the Euclidean amplitude 
\cite{Cornalba:2007zb}.
The Regge regime can be reached starting from the  Euclidean four-point function, rotating $z$ anti-clockwise around the branch points at 0 and 1, keeping $\bar{z}$ fixed, and then considering the limit $z,\bar{z} \to 0$.

We wish to determine the Regge limit of the four-point function associated to the bulk interaction
\ba
\phi^2 (\nabla^2)^k \phi^2= 2^k \phi^2 (\nabla_{\mu_1} \dots \nabla_{\mu_k} \phi)^2 +\dots\ ,
\ea
where the $\dots$ stand for terms with fewer derivatives after using the equations of motion.
We shall restrict to the case of even $k$ corresponding to the left most box in each row in Fig.~\ref{tabela}.
The exact computation of the four-point function could be performed using the techniques of the previous section to reduce the Witten diagrams,
\ba
2^k \sum_{{\rm perm\ } P_{i}}\int_{\rm AdS} &&dX\, (-2P_2\cdot X)^{-\Delta} (-2P_3\cdot X)^{-\Delta} \label{del2m}\\
&&\nabla_{M_1}\dots \nabla_{M_k}  (-2P_1\cdot X)^{-\Delta}\nabla^{M_1} \dots  \nabla^{M_k} (-2P_4\cdot X)^{-\Delta} \ ,\nonumber
\ea
to a sum of D-functions.
However, one can determine the Regge limit of the four-point function directly.
The basic idea is that, in this limit, the integral over the interaction point in AdS is dominated by the points null related to the external points $P_1$  and $P_4$. Notice that these points are also almost null related to  $P_3$  and $P_2$ because $P_3 \sim -P_1$  and $P_2 \sim -P_4$.
More precisely, we can introduce coordinates in AdS via
\ba
X=\left(u,v\left(1-\frac{uv}{4\cosh^2 r}\right),\cosh r\left(1-\frac{uv}{2\cosh^2 r}\right),e \sinh r \right) \ ,
\ea
with $e\in S^{d-2}$.
In the region of interest we have $uv\ll \cosh^2 r$ and we can simply write
\ba
X\approx (u,v,w)
\ea
with $w\in H_{d-1}$. The $u=v=0$ hypersurface is the intersection of the lightcones of $P_1$ and $P_4$, as shown in
 Fig. \ref{ReggeAdS}.
We can now determine the Regge limit of a given Witten diagram using the following simple rules
\ba
\int_{\rm AdS} dX \dots &\approx&\int  \frac{du\,dv}{2} \int_{H_{d-1} }dw \dots\ ,\\
-2X\cdot P_1&\approx& v \ ,\\
-2X\cdot P_2&\approx&u - 2\bar{x}\cdot w\ , \\
-2X\cdot P_3&\approx&-v -2 x\cdot w\ ,\\
-2X\cdot P_4&\approx&-u\ .
\ea
The Regge limit of (\ref{del2m}) is then given by
\be
\int  \frac{du\,dv}{2}\int_{H_{d-1} }dw \frac{4i  (g^{uv})^k  2^k}{(u-2\bar{x}\cdot w +i\epsilon)^\Delta(-v-2x\cdot w +i\epsilon)^\Delta}
\partial_u^k\frac{1}{(-u +i\epsilon)^\Delta}\partial_v^k\frac{1}{(v+i\epsilon)^\Delta} \ ,
\ee
where we have used the fact that the dominant behavior is obtained by taking the maximum number of $u$ and $v$ derivatives.
The factor of 4 comes from the 4 possible permutations of the external $P_i$ that give the same dominant behavior and the factor of $i$ comes from the Wick rotation.
The integrals over $u$ and $v$ can be done to give
\ba
i \pi^2 2^{2k+1}\frac{\Gamma^2(2\Delta+k-1)}{\Gamma^4(\Delta)}\int_{H_{d-1} }  dw\,
\frac{1}{(-2x\cdot w +i\epsilon)^{2\Delta+k-1}(-2\bar{x}\cdot w +i\epsilon)^{2\Delta+k-1}} \ .
\ea
As explained in \cite{Cornalba:2007fs}, this convolution integral over $H_{d-1} $ can be evaluated using harmonic analysis on hyperbolic space.
This leads to the final expression
\ba
(z\bar{z})^\Delta \mathcal{A}_1 (z,\bar{z}) \approx i  \sigma^{1-k}  \int d\nu \,
\frac{ \pi^d \Gamma^2\left(\frac{2\Delta+k+i\nu-d/2}{2}\right)  \Gamma^2\left(\frac{2\Delta+k-i\nu-d/2}{2}\right)}{2^{1-2k}\Gamma^4(\Delta)}\,
 \Omega_{i\nu} (\rho) \ , \label{ReggeL}
\ea
where $z=\sigma e^{\rho}$, $\bar{z}=\sigma e^{-\rho}$ and $ \Omega_{i\nu}$ are harmonic functions on $H_{d-1} $.
In $d=2$, the harmonic functions are simply cosines,
\ba
\Omega_{i\nu} (\rho)=\frac{1}{2\pi}\cos(\nu \rho) \ ,
\ea
and in $d=4$, they are given by
\ba
\Omega_{i\nu} (\rho)= \frac{\sin(\nu \rho)}{4\pi^2 \sinh \rho} \ .
\ea

Now that we have determined the Regge limit of the four-point function we can study its conformal partial wave decomposition.
We shall follow closely the methods of \cite{Cornalba:2007fs}. In particular, our starting point will be the representation
\ba
(z\bar{z})^\Delta \mathcal{A}_1(z,\bar{z})=\sum_{l=0}^L \int d\nu\, f_l(\nu) \mathcal{G}_{i\nu,l} (z,\bar{z})\ ,
\ea
where $f_l(\nu) =f_l(-\nu) $ and
\ba
\mathcal{G}_{i\nu,l} (z,\bar{z}) = t(\nu,l) g_{\frac{d}{2}+i\nu,l} (z,\bar{z})+t(-\nu,l) g_{\frac{d}{2}-i\nu,l} (z,\bar{z})\ ,
\ea
with
\ba
t(\nu,l)=-\frac{ \Gamma\left(i\nu+\frac{d}{2}-1\right)\Gamma^4\left(\frac{l+i\nu }{2}+\frac{d}{4}\right)}
{4\pi^{\frac{d}{2}}\Gamma(i\nu) \Gamma\left(l+i\nu+\frac{d}{2}\right)\Gamma\left(l+i\nu+\frac{d}{2}-1\right)}\ .
\ea
As we shall see, the standard conformal partial wave decomposition can be obtained from this representation by reducing 
the $\nu$ integral to a sum of the residues of the poles along the $\nu$ imaginary axis.
This representation is particularly convenient because in the Regge limit we have 
\ba
\mathcal{G}_{i\nu,l} (z,\bar{z}) \approx 2\pi i \,\sigma^{1-l} \,\Omega_{i\nu} (\rho)\ .
\ea
Therefore, the Regge limit of the full amplitude is determined by the   highest spin function $f_L(\nu)$,
\ba
(z\bar{z})^\Delta \mathcal{A}_1(z,\bar{z})\approx 2\pi i \,\sigma^{1-L} \, \int d\nu f_L(\nu) \Omega_{i\nu} (\rho)\ .
\ea
Comparing with (\ref{ReggeL}) we conclude that $k=L$ and
\ba
f_L(\nu) =   \frac{ \pi^{d-1} 2^{2L-2}}{\Gamma^4(\Delta)}
\Gamma^2\left(\frac{2\Delta+L+i\nu-d/2}{2}\right)  \Gamma^2\left(\frac{2\Delta+L-i\nu-d/2}{2}\right)\ .
\ea
In order to recover the standard conformal partial wave expansion we use the parity of $f_l(\nu)$ to write
\ba
(z\bar{z})^\Delta \mathcal{A}_1(z,\bar{z})=2 \sum_{l=0}^L \int d\nu f_l(\nu) t(\nu,l) g_{\frac{d}{2}+i\nu,l} (z,\bar{z})\ .
\ea
We can now deform the $\nu$ integration contour into the lower half plane and pick up the poles of the integrand.
The function $t(\nu,l)$ does not have poles in the lower half plane 
and the poles of $g_{\frac{d}{2}+i\nu,l}$ give rise to partial waves with spin smaller than $l$.
The function $ f_L(\nu)$ has double poles at $\frac{d}{2}+i\nu=2\Delta+2n+L$ for $n=0,1,2,\dots $.
It is convenient to define
\ba
\nu(n)=- i \left( 2\Delta+2n+L-\frac{d}{2}\right) \ ,
\ea
and write
\ba
f_L(\nu(n)) =    \frac{ \pi^{d+1} 2^{2L-2}}{\Gamma^4(\Delta)}
\frac{\Gamma^2\left(2\Delta+L+n-d/2 \right) }{ \Gamma^2\left(1+n \right) \sin^2(\pi n)} \ .
\ea
Then,
\be
(z\bar{z})^\Delta \mathcal{A}_1(z,\bar{z})=  - \sum_{n=0}^\infty \frac{\partial}{\partial n}
\left( \frac{  \pi^{d} 2^{2L+1}\Gamma^2\left(2\Delta+L+n-d/2 \right)  
t\left(\nu(n),L\right)}{\Gamma^4(\Delta)\Gamma^2\left(1+n \right) }
g_{2\Delta+2n+L,L} (z,\bar{z}) \right)+\dots\ ,
\ee
where the $\dots$ denote partial waves with spin smaller than $L$.
Comparing with (\ref{A1cross}) we conclude that 
\ba
p_1(n,L)=\frac{\partial}{2\partial n}\left( p_0(n,L)\gamma(n,L)\right)\ ,
\ea
in agreement with (\ref{cdg}) and
\ba
p_0(n,L)\gamma(n,L)&=& \frac{ \pi^{\frac{d}{2}} 2^{2L} \Gamma\left( 2\Delta+2n+L-1\right)}
{\Gamma^4(\Delta)\Gamma^2\left(1+n \right)\Gamma\left(2\Delta+2n+L-d/2 \right)}\\&\times&
\frac{ \Gamma^2\left(2\Delta+n+L-d/2 \right)
\Gamma^4\left(\Delta+n+L \right) }
{ \Gamma\left(2\Delta+2n+2L\right)
\Gamma\left(2\Delta+2n+2L-1\right)}\ . \nonumber
\ea
Using the explicit expression (\ref{3pt2D}) for $p_0(n,l)$ in $d=2$, we find
\ba
\gamma(n,L)= \frac{ \pi \Gamma(n+L+1)\Gamma\left(2\Delta+n+L-1\right)
\Gamma\left(\Delta+n-\frac{1}{2} \right)\Gamma\left(\Delta+n+L \right) }
{4\Gamma\left(1+n \right)
\Gamma\left(\Delta+n \right)\Gamma\left(\Delta+n+L+\frac{1}{2} \right)\Gamma\left(2\Delta+n-1\right)}\ .
\ea
This is precisely the solution (\ref{gammaL}) found in the previous section.
In $d=4$ we find, using (\ref{p0d4}),
\begin{equation}
\gamma(n,L)=\frac{\pi^2\Gamma(n+L+2)\Gamma(\Delta+n-\frac{3}{2})\Gamma(\Delta+n+L)\Gamma(2\Delta+n+L-2)}{16(1+L)(\Delta-1)^2\Gamma(n+1)\Gamma(\Delta+n-1)\Gamma(\Delta+n+L+\frac{1}{2})\Gamma(n+2\Delta-3)}
\end{equation}
in agreement with (\ref{d4maximalspingamma}).

\sect{Locality and the CFT singularity}
\label{localizing}

\begin{figure} 
 \centering
\includegraphics[width=8cm]{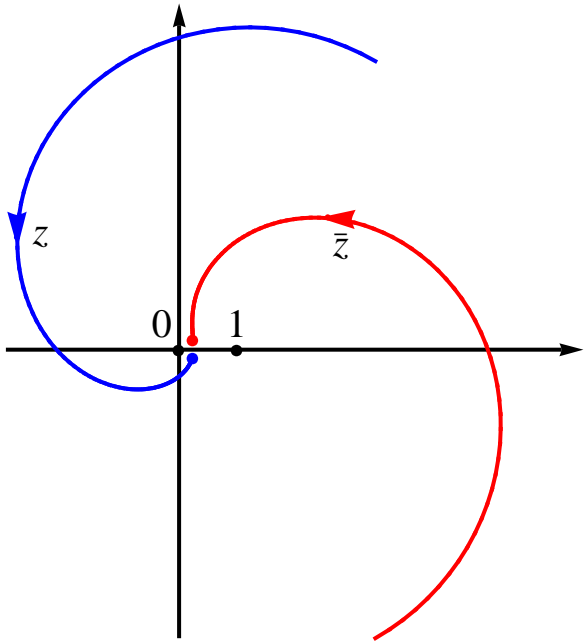}
\caption{ 
Complex paths of $z$ and $\bar{z}$ from the Euclidean regime to the Lorentzian  regime.
The path of $\bar{z}$ is equivalent to going around the branch points at 0 and 1 anticlockwise and then following the complex conjugate of the $z$-path.
The final Lorentzian values are given by $z,\bar{z}=e^{\pm \rho} \sin^{2}\frac{\theta}{2}   \mp i\epsilon$.}
\label{Wickrotation}
\end{figure}

As reviewed in section \ref{AdSscat} the existence of a local bulk theory implies a peculiar behavior of the CFT correlators.
In \cite{Gary:2009ae}  the scattering thought experiment described in section \ref{AdSscat} was used to predict
a particular singularity of the CFT four-point function.
In the present notation this prediction reads
\be
(z\bar{z})^\Delta \mathcal{A}_1(z,\bar{z}) \approx \frac{\mathcal{F}(\sigma)}{\rho^{2\beta}} \ , 
\ \ \ \ \ \ (\rho \to 0)\ ,\label{singu}
\ee
where we recall the relations $z=\sigma e^\rho$ and  $\bar{z}=\sigma e^{-\rho}$.
It is crucial that the limit $\rho \to 0$ is taken after analytically continuing the Euclidean correlator 
to the Lorentzian regime of the scattering process.
More precisely, this analytic continuation corresponds to 
the Wick rotation of AdS global time $\tau \to -i \tau e^{i\alpha}$ where $\alpha=0$ is the Euclidean regime and 
$\alpha=\frac{\pi}{2}$ is the Lorentzian one.
For $\rho=0$, this gives the following continuation of the cross ratios  \cite{Gary:2009ae}
\ba
z(\alpha)=\cos^2\frac{\theta-i\pi e^{i\alpha}}{2}\ ,\ \ \ \ \ \ \ \ \ \ 
\bar{z}(\alpha)=\cos^2\frac{\theta+i\pi e^{i\alpha}}{2}\ ,
\ea
shown in Fig.~\ref{Wickrotation}. 
The strength of the singularity is controlled by 
\be
2\beta=4\Delta+2k-3\ ,
\ee
where $2k$ is the number of derivatives in the quartic vertex of our scalar model.
More generally, the strength of the singularity is fixed by the scaling dimension of the bulk interaction vertices involved
in the computation of the correlator.
Finally, the residue of the singularity is directly related to the S-matrix of the bulk theory, 
\be
T(s,t) \propto s^{k}  \frac{ \mathcal{F}(\sigma)} { \sigma^{1-k} (1-\sigma)^{2\Delta+k-2}}\ ,
\ee
where $\sigma$ is related to the Mandelstam invariants $s$ and $t$ and the scattering angle $\theta$ via
\be
\sigma=-\frac{t}{s}=\sin^2\frac{\theta}{2}\ .
\ee
Notice that after the flat space limit, $R \to \infty$, the external particles have zero mass, $m^2=\Delta(\Delta-d)/R^2 \to 0$.
Therefore, the three Mandelstam invariants obey $s+t+u=0$.

We now wish to understand the origin of this singularity from the point of view of the conformal partial wave expansion.
In order to recover the usual s-channel partial wave expansion in flat space we consider the conformal partial wave expansion in the same channel. 
The first step is to study the Lorentzian $\rho \to 0$ limit of a single conformal partial wave $g_{E,l}(1/z,1/\bar{z})$.
It is clear from the explicit expressions (\ref{CPW2d}) and  (\ref{CPW4d}) that, in $d=2$, the partial waves 
are not singular in this limit and that, in $d=4$, they have a $1/\rho$ singularity. 
We see that, in general, the conformal partial waves have a weaker singularity than the full four-point function.\footnote{The unitarity bound in $d=4$ requires $\Delta \ge 1$ and therefore the four-point function is 
always more singular than a single partial wave. In $d=2$ we restrict ourselves in the present discussion to the case $\beta > 0$.}
Therefore, the singularity must emerge from the infinite sum over conformal dimensions $E$,
and so it is sufficient to consider the large $E$ behavior of the analytically continued partial waves.
This can be easily obtained by a saddle point approximation.

First, we consider the large $\alpha$ behavior of the basic function
\be
k(2\alpha,1/z)=z^{-\alpha} F(\alpha,\alpha,2\alpha,1/z)= \frac{\Gamma(2\alpha)}{\Gamma^2(\alpha)}
\int_0^1\frac{dt}{t(1-t)} \left(\frac{t(1-t)}{z-t}\right)^\alpha\ .
\ee
There are two saddle points in the complex $t$-plane,
\be
t_\pm=z\pm\sqrt{z^2-z}\ ,
\ee
and the saddle point approximation gives
\be
k(2\alpha,1/z)\approx \frac{2^{2\alpha-1}}{\sqrt{t(1-t)}} (2t-1)^\alpha\ .\label{saddlek}
\ee
For $z<1$ and real the two saddle points $t_\pm$ are the
complex conjugate of each other, and the integrand has a branch point at $z$ between 0 and 1.
To define the integral we need an $i\epsilon$ prescription to move the branch point away from the integration contour.
This prescription is given by the analytic continuation shown in Fig.~\ref{Wickrotation},
\be
z,\bar{z}=e^{\pm\rho}\sin^{2}\frac{\theta}{2} \mp i\epsilon\ .
\ee
Therefore, for $z$ we pick up the contribution from the saddle point $t\equiv t_+(z)$ with positive imaginary part and for $\bar{z}$ 
we pick up the contribution from the saddle point $\bar{t}\equiv t_-(\bar{z})$ with negative imaginary part.
Expanding at small $\rho$ we obtain
\be
t,\bar{t}=\pm i  e^{\mp i \frac{\theta}{2} }\sin \frac{\theta}{2} +\frac{i}{2} \rho  e^{\mp i \theta } \tan \frac{\theta}{2} +\mathcal{O}(\rho^2)\ .
\ee
Knowing the asymptotic behavior of the basic function   $k(2\alpha,z)$ we are ready to determine the large $E$ and small $\rho$
behavior of the Lorentzian partial waves.

We start by the $d=2$ case.
Using (\ref{saddlek}) it is easy to obtain
\be
g_{E,l}(1/z,1/\bar{z})\approx \frac{e^{-i\pi E} 2^{2E-3} }{ \sin \theta} e^{-i E\rho \tan\frac{\theta}{2}}
P_l^{(2)}(\theta)\ ,
\ee
where
\be
P_l^{(2)}(\theta)=\frac{8\cos(l\theta)}{1+\delta_{l,0}}\ .
\ee
We denote with $P_l^{(d)}(\theta)$ the harmonic functions on $S^{d-1}$ with laplacian eigenvalue $-l(l+d-2)$.
A convenient normalization is
\be
P_l^{(d)}(\theta)=
\frac{2^d \pi ^{\frac{d-1}{2}} (d+2 l-2) \Gamma (d+l-2)   }{\Gamma  \left(\frac{d-1}{2}\right) \Gamma (l+1)}
 F\left(-l,d+l-2,\frac{d-1}{2},\sin^2 \frac{\theta }{2}\right)\ ,
\ee
such that $T(s,t)=s^{\frac{3-d}{2}}\sum_{l=0}^\infty P_l^{(d)}(\theta)$ corresponds to free propagation 
\cite{Penedones:2007ns}.
The s-channel conformal partial wave decomposition of the four-point function is 
\be
 (z\bar{z})^\Delta {\cal A}_1(z,\bar z)  = 
\sum_{n=0}^\infty \sum_{l=0 \atop \rm even}^L \frac{\partial}{2\partial n} \Big( \gamma (n,l) p_0(n,l)g_{2\Delta+2n+l,l}(1/z,1/\bar{z})
\Big) \ .
\ee
Using the large $n$ approximation
\begin{equation}
 p_0(n,l) \approx \frac{ \pi }{\Gamma^4(\Delta)} \frac{n^{4\Delta-3}}{2^{4\Delta+4n+2l-5}} \ ,
\end{equation}
we conclude that in order to recover the right singularity (\ref{singu}) we need 
\be
\gamma(n,l)\approx \mu_l n^{2k-1}\ ,
\ee
at large $n$. Then,
\ba
  (z\bar{z})^\Delta{\cal A}_1(z,\bar z) 
\approx -
  \frac{ 4i \pi^2 }{\Gamma^4(\Delta)}  
\frac{\left(\sin\frac{\theta}{2}\right)^{4\Delta}}{ \sin \theta} 
  \sum_{l=0 \atop \rm even}^L \mu_l P_l^{(2)}(\theta)
 \sum_{n=0}^\infty
n^{4\Delta+2k-4} 
e^{-2 i n \rho \tan\frac{\theta}{2}} \ .
\ea
The small $\rho$ behavior of the sum over $n$ can be determined by approximating the sum by an integral.
This gives the predicted singularity,
\ba
  (z\bar{z})^\Delta {\cal A}_1(z,\bar z) \approx
  \frac{ \pi^2 \Gamma(4\Delta+2k-3)}{4\Gamma^4(\Delta)(2i)^{4\Delta+2k-6}}  
\frac{\left(\sin \frac{\theta}{2}\right)^{2-2k} \left(\cos \frac{\theta}{2}\right)^{4\Delta+2k-4}   }{\rho^{4\Delta+2k-3}} 
  \sum_{l=0 \atop \rm even}^L \mu_l P_l^{(2)}(\theta)
\ea
and the standard partial wave expansion of the flat space S-matrix,
\be
T(s,t) \propto s^{k} \sum_{l=0 \atop \rm even}^L \mu_l  P_l^{(2)}(\theta)\ .
\ee
Notice that the condition $k \ge L$ guarantees that $T$ is always a polynomial of $s$ and $t$.
In the simplest example of a pure quartic interaction with $k=L=0$ we obtain an S-matrix independent on the 
Mandelstam invariants, as expected.
In case $L=2$ we obtain the simple expression
\be
T(s,t)\propto s^{k} (\mu_0 +2 \mu_2 \cos(2\theta))\ .
\ee
In appendix \ref{expCPW} we present the exact conformal partial wave decomposition of two $L=2$ examples, 
one with $k=2$ and one with $k=3$.
The first example is the interaction $(\nabla \phi)^2 (\nabla \phi)^2 $ which has $k=2$ and $\mu_2/\mu_0=7/2$.
This gives $T(s,t)\propto s^2+ts+t^2\propto s^2+t^2+u^2 $, as expected.
The second example is the interaction $(\nabla \phi)^2 (\nabla_\mu \nabla_\nu \phi)^2 $ which has $k=3$ and $\mu_2/\mu_0=-1/2$.
This yields again the expected result, 
$T(s,t)\propto st(s+t)\propto s^3+t^3+u^3 $.

Similarly, in $d=4$ we find that 
\be
g_{E,l}(1/z,1/\bar{z})\approx-i \frac{e^{-i\pi E} 2^{2E-9} }{\pi (1+l) \rho \sin^2 \frac{\theta}{2}} e^{-i E\rho \tan\frac{\theta}{2}}
P_l^{(4)}(\theta)\ ,
\ee
where
\be
P_l^{(4)}(\theta)=64\pi(1+l)\frac{\sin(l+1)\theta}{\sin\theta}\ 
\ee
are harmonic functions on $S^3$.
Using the large $n$ limit of (\ref{p0d4}),
\begin{equation}
 p_0(n,l) \approx \frac{ \pi (l+1)}{\Gamma^2(\Delta)\Gamma^2(\Delta-1)} \frac{n^{4\Delta-6}}{2^{4\Delta+4n+2l-7}}  
\end{equation}
and requiring the singularity (\ref{singu}), we obtain the large $n$ behavior of the anomalous dimensions,
\begin{equation}
 \gamma (n,l) \approx \mu_l n^{2k+1}
\, .
\end{equation}
This gives 
\ba
  (z\bar{z})^\Delta{\cal A}_1(z,\bar z) \approx 
  \frac{ \pi(\Delta-1)^2 \Gamma(4\Delta+2k-4)}{16 \Gamma^4(\Delta)(2i)^{4\Delta+2k-6}}  
\frac{\left(\sin \frac{\theta}{2}\right)^{2-2k} \left(\cos \frac{\theta}{2}\right)^{4\Delta+2k-4}   }{\rho^{4\Delta+2k-3}} 
  \sum_{l=0 \atop \rm even}^L \mu_l  P_l^{(4)}(\theta)\ 
\ea
and we recover the partial wave decomposition of the flat space S-matrix,
\be
T(s,t) \propto s^{k}  \sum_{l=0 \atop \rm even}^L \mu_l  P_l^{(4)}(\theta)\ .
\ee

\sect{Convergence in $l$}

The story thus far fits together nicely, but there is a significant loose end.   We introduced the maximum spin $L$ as a device to allow counting of solutions, but we needed to use this rather early in the process of solving the constraints, beginning in Eq.~(\ref{ln1zb}).  Thus there is a concern that we might be missing some nonlocal solutions involving unbounded $l$.  Here we analyze this issue, and largely exclude it.

First we need to understand how crossing constrains the large-$l$ behavior of $\gamma(n,l)$.  Write
\begin{equation}
{\cal A}_1(z,\bar z) = \alpha(z,\bar z) \ln z\bar z + \beta(z,\bar z)\ , \label{albet}
\end{equation}
where $\alpha(z,\bar z)$ and $\beta(z,\bar z)$, regarded as independent functions of $z$ and $\bar z$, are holomorphic around $z=0$.  From the OPE expansion it follows that their only singularities in $z$ or $\bar z$ are branch cuts from $1$ to $\infty$.  The amplitude grows as $\ln( 1 - \bar z)$ as $\bar z \to 1$, and we assume that this is true of the individual functions $\alpha(z,\bar z)$, $\beta(z,\bar z)$.\footnote{It would seem impossible for more singular terms to cancel in Eq.~(\ref{albet}), because one term has a discontinuity and the other not, but we do not have a derivation.}

Then for $l > n$,
\begin{equation}
\gamma(n,l) = 
\oint_C \frac{dz}{2\pi i z^{n+1}}
\oint_C \frac{d\bar z}{2\pi i \bar z^{n+l+1}}   F_{1 -\Delta - n}(z) F_{1 -\Delta - n - l}(\bar z) \alpha(z,\bar z)\ ,
\end{equation}
where both contours circle the origin.
For large $l$ we can estimate this by expanding the $\bar z$ contour, the factor of $\bar z^{-(n+l+1)}$ pushing the contour outward until it wraps the branch cut at $1$, and the dominant contribution comes from the neighborhood of this point.  We then use the singular behavior $\ln(1-\bar z)$ known from crossing to conclude that the integral is no larger than $O(l^{-1} \ln l)$; the precise bound does not matter because we can readily improve it. 
Note as in Eq.~(\ref{eigen}) that 
\begin{equation}
\bar {\cal D} F_{1 -\Delta - n - l}(\bar z) = \lambda_{\Delta + n + l} F_{1 -\Delta - n - l}(\bar z)\ , \quad \lambda_{\Delta + n + l} = (\Delta + n + l) (\Delta + n + l - 1) \ .
\end{equation}
We can multiply $F_{1 -\Delta - n - l}(\bar z)$ in the contour integral by $1 = \bar {\cal D} /\lambda_{\Delta + n + l}$ and then integrate by parts to have $\bar {\cal D}$ act on $\alpha(z,\bar z)$.  Now, $\bar {\cal D}$ has the convenient property that when acting on $\alpha(z,\bar z)$, the singularity at $\bar z = 1$ is still 
$\ln(1-\bar z)$; to see this consider its action on a general monomial $(1-\bar z)^m \ln(1-\bar z)$.  Thus we improve the bound by a factor of $\lambda_{\Delta + n + l}^{-1} = O(l^{-2})$.  By iterating we can conclude that at fixed $n$, $\gamma(n,l)$ must fall faster than any power of $l$,
\begin{equation}
l^m \gamma(n,l) \to 0 \ {\rm as}\ l \to \infty\ ,\ {\rm\ all}\ m, n\ . \label{falloff}
\end{equation}

We can now extend the upper limits on the $l$ sums~(\ref{ln1zb}, \ref{lnln}, \ref{pq}) to infinity.  The sum over $l$ converges sufficiently rapidly that it cannot generate additional singularities of the form of $(1-\bar z)^m \ln(1-\bar z)$ so these come only from the explicit logarithms in the hypergeometric functions.  In particular, the bound on $\gamma(n,l)$ together with the asymptotic behavior of the $J(p,q)$ implies that the sums~(\ref{pq}) converge when extended to infinity.  Thus, for example, we can immediately use this relation at $(p,q) = (1,0)$ to obtain 
\begin{equation}
\gamma(1,0) = -\frac{1}{J(1,0)} \sum_{l=2 \atop \rm even}^\infty  \gamma(1,l) J(1+l,0) + \frac{1}{J(1,0)}\sum_{l=0 \atop \rm even}^\infty  \gamma(0,l) J(l,1) \ . \label{g00}
\end{equation}
Similarly, at each $p$ we can use the equations with $q< p$ and the invertibility of $M(p)_{qr}$  to solve for $\gamma(p,0), \ldots, \gamma(p, 2p-2)$.  In all, the $\gamma(n,l)$ with $l \geq 2n$ are free parameters, and we solve for $\gamma(n,l)$ with $l \leq 2n -2$, just as in the earlier discussion where we had the additional condition that $l \leq L$.\footnote{The redundancy of the constraints noted earlier shows up here as the result, following from the known bulk solutions, that if the free parameters $\gamma(n,l \geq 2n)$ vanish for $l > L$, then so do the $\gamma(n,l < 2n)$ for $l > L$.}  

Eq.~(\ref{g00}), and together with the higher-$p$ equations, give all $\gamma(n,l)$ as convergent sums of fixed-$L$ solutions, one for each free parameter.  Thus the solutions found earlier are complete.

There would seem to be a trivial counterexample to our conjecture.  Since we have an infinite number of higher-derivative solutions in the bulk, it would seem that we can make a nonlocal solution by taking an infinite sum.  Of course we expect such nonlocality on the scale $l_{\rm s}$, from integrating out the string-scale and higher states.  General arguments from effective field theory restrict the coefficients of higher-derivative operators to be set by the cutoff scale, in order that that net positive powers of the cutoff not appear in loops.  In effective field theory one works to a given order in the inverse cutoff scale, and then only a finite number of higher-derivative terms can appear.  
We would expect this argument to have a parallel in the CFT: going to higher orders in $1/N^2$ we will encounter divergences in the sum over intermediate states in the low-dimension sector that we are studying.  These will be cut off in the full theory, and existence of the $\Delta_{\rm large} \to \infty$ limit should require that their coefficients scale as inverse powers of  $\Delta_{\rm large} $, such that 
there will be only a finite number of solutions to any given order.

It is worth exploring a bit further the possibility of constructing a non-local bulk interaction of the form
\ba
\sum_{k=0}^\infty d_k \, \phi^2 (a^2 \nabla^2 )^k \phi^2\ ,
\ea
where $d_k$ are dimensionless coefficients and $a$ is a characteristic length scale of the interaction.
The partial wave expansion is
\be
\gamma(n,l) = \sum_{k=0}^\infty d_k \gamma^{(k)}(n,l)\ , \label{sumbulk}
\ee
where the $\gamma^{(k)}(n,l)$ corresponding to these operators were partly obtained in Sec.~5.2.
Based on the coefficients obtained in Sec.~5.2, 
\begin{equation}
\gamma^{(k)}(n,k) \sim (k!)^2 k^{2n}\ .
\end{equation}
We do not have an explicit solution for partial waves $l \neq k$, but can estimate $\gamma(n,l)$ by keeping only $k=l$ on the right, 
\be
\gamma(n,l) \sim  d_l (l!)^2 l^{2n}\ . \label{sumop}
\ee

Consider the example $d_k =  (a/R)^{2k}$, which is just what would be obtained from integrating out a particle of mass $a^{-1}$.  The estimate~(\ref{sumop}) is inconsistent with the falloff~(\ref{falloff}), so this does not solve the crossing relation.  The reason is clear: the particle that we have integrated out corresponds to a new single-trace operator with $\Delta(\Delta-d) = R^2/a^2$, and this must be included explicitly in the OPE.  Even a gaussian nonlocality, $d_k\sim (a/R)^{2k}/k!$, fails to satisfy crossing.  We need the much stronger condition that $(k!)^2 d_k$ falls faster than any power of $k$.  This condition is reminiscent of the generalized notion of locality in Ref.~\cite{Jaffe:1967nb}, and suggests that even with an infinite series of local operators the crossing condition imposes some form of locality.\footnote{We thank D. Gross for informing us of Ref.~\cite{Jaffe:1967nb}, and for discussions.}

\sect{Inclusion of $T_{\mu\nu}$}

Let us first consider the effect of dropping the ${ \mathbb{Z}}_2$ symmetry.  The operator ${\cal O}$ will
generically appear in the product ${\cal O}{\cal O}$, leading to an extra term
\begin{equation}
c_{\cal OOO}^2
 \frac{g_{\Delta,0} (z,\bar{z})}{(z\bar z)^{\Delta}}
 \end{equation}
in the partial wave expansion~(\ref{CPW}).  At fixed $c_{\cal OOO}$ we can think of this as an inhomogeneous term in the crossing relation~(\ref{crossing}, \ref{pq}, \ref{crossingeqns4d}).  Any two solutions will differ by a solution to the homogenous equation, as already studied, so the effect is to introduce at most one new parameter $c_{\cal OOO}^2$ into the solution.  As before, the bulk picture provides an existence proof for these solutions, generated by the exchange graph with two $\phi^3$ couplings, so the counting in the bulk and the CFT again matches.  We will explore the detailed form of these solutions in future work.  We note that they will necessarily involve all values of $l$.

Similarly we can add additional scalar operators of various dimensions into the OPE.  Further we can spin-2 operators of various dimensions, where $\Delta = d$ would be the energy momentum tensor.  Each additional operator introduces one additional parameter into the crossing solution, and correspondingly one bulk coupling.  Thus we can immediately embed our result into a full-fledged \mbox{CFT}.  In fact, at leading order in the planar expansion our result would apply to the correlator of four identical scalar operators in {\it any} \mbox{CFT}.  Thus we can conclude that our conjecture holds, to this order, quite generally.

The same logic would allow us to introduce operators of spin greater than two, apparently giving a result even more general than we conjectured.  From the bulk point of view, we are adding a traceless symmetric field $\phi_{\mu_1\ldots\mu_l}$.  The flat-space propagator for mass $M^2$ would be 
\begin{equation}
\langle \phi_{\mu_1\ldots\mu_l}\phi_{\nu_1\ldots\nu_l} \rangle = \frac{1}{k^2 + M^2} S P_{\mu_1\nu_1} \ldots P_{\mu_l\nu_l}\ ,
\end{equation}
where
\begin{equation}
P_{\mu\nu} = \eta_{\mu\nu} + \frac{k_\mu k_\nu}{M^2}
\end{equation}
and $S$ projects onto the symmetric traceless part.  In AdS spacetime this is readily made covariant. 
The operator $P_{\mu\nu}$ removes ghosts from the timelike components, as is clear in the rest frame.  The inverse power of $M^2$ implies that loops will contain positive powers of the cutoff scale; these would be absent only if $\phi_{\mu_1\ldots\mu_l}$ coupled to a higher-spin conserved current, which is not available in an interacting theory.  Thus such fields cannot be present in the low energy effective theory.\footnote{Such arguments have recently been explored in Ref.~\cite{Porrati:2008ha}.}  As in the previous section, we would expect such bulk arguments to be reflected in a breakdown of the $1/N^2$ expansion in the CFT.

\sect{Conclusions}

We have confirmed our locality conjecture in the case of the correlator of four low-dimension operators to leading nontrivial order in $1/N^2$.  This excludes the possibility that the bulk theory is somehow smeared over the AdS scale, and so closes a potential loophole in AdS/CFT duality.  In particular, a mysterious property of the four-point function is shown to follow from a simple property of the operator spectrum.  

Our analysis was limited to CFTs in $d=2$ and in $d=4$. The extension to $d=6$ should be relatively straightforward using the known explicit expressions for conformal partial waves \cite{Dolan:2000ut}. On the other hand, the extension to the $d=3$ case, relevant for condensed matter applications, can not be done using the techniques of this paper because the simplest known form of the conformal partial waves in odd dimension are the integral representations of \cite{classic}.
We are presently trying to generalize our methods so that they do not rely so heavily on the knowledge of explicit expressions of the conformal partial waves and can thus be valid in any dimension.

One could consider the extension to higher orders in $1/N^2$; we have noted some potential obstructions, from effective field theory reasoning.  Also of interest is the coupling of stringy states to the low dimension sector that we have studied.   More far-reaching directions would include scattering of gravitons (that is, correlators of $T_{\mu\nu}$) and their relation to black hole physics.\footnote{J. de Boer and collaborators have been considering these subjects independently.}   
Further, there are many examples of gauge/gravity duality without conformal invariance.  Although conformal invariance has played a major role in our work, there should be a nonconformal extension though it will have many more equations and unknowns.  

The general direction of our work is to give a derivation of the low energy sector of gauge/gravity duality from the assumptions of a large-$N$ expansion and a gap in the dimensions, without an explicit string construction.  If we assume a certain spectrum of low-dimension operators, the bulk dual necessarily follows.  This provides a context for extending the range of AdS/CFT duality, and is likely to be useful in applications to condensed matter physics and in the study of cosmological spacetimes.\footnote{As a curious application, suppose that we have a CFT whose only low-dimension operators are the energy-momentum and a conserved current.  Then at finite charge density, such a system necessarily violates the third law of thermodynamics.  The point is that --- assuming that our result for the four-point function applies to the partition function as well --- then the latter is given by the Reissner-Nordstrom black hole entropy, whose horizon area is finite even at zero temperature.}  Of course, there is no guarantee that a given low-dimension spectrum can be embedded in a full CFT, and all known examples arise from string backgrounds.  

\subsection*{Acknowledgments}

We have benefited from discussions with many colleagues, including O. Aharony, T. Banks, J, de Boer, D. Berenstein, L. Cornalba, M. Costa, M. Douglas, M. Gary, S. Giddings, D. Gross, S. Hartnoll, G. Horowitz, J. McGreevy, T. Okuda, S. Shenker, E. Silverstein, E. Sokatchev, M. Strassler, L. Susskind, S. Todadri, and P. Vieira, and with many participants at the KITP Program on Fundamental Aspects of String Theory.  
 JP$_1$ is funded by the FCT fellowship SFRH/BPD/34052/2006 and partially funded by the grant CERN/FP/83508/2008.
This work was supported in part by NSF grants PHY05-51164 and PHY07-57035. 

\begin{appendix}

\sect{Some properties of $J(p,q)$}
\label{J}

The integral $J(p,q)$ defined in equation (\ref{defJ}) can be explicitly performed in terms of a sum over hypergeometric functions $\,_4F_3$ at $z=1$,
\begin{align}
 J(p,q) &=-\frac{C_p}{C_q}\frac{\Gamma(2\Delta+2p)}{\Gamma^2(\Delta+p)}\sum_{l=0}^{p} {p\choose l} \frac{(-1)^l(\Delta+p)_{q-l}^2}{\Gamma^2(q+1-l)}
\nonumber\\ &\times
\,_4F_3(-q+l,-q+l,1-\Delta-q,1-\Delta-q;\label{a1}
\\\nonumber
&\qquad\qquad\qquad\qquad\qquad\qquad
l-\Delta-q-p+1,l-\Delta-q-p+1,2-2\Delta-2q;1)\ . 
\end{align}
where $(a)_b = \Gamma(a+b)/\Gamma(a) = a(a+1)\ldots(a+b-1)$.
For a few cases this expression simplifies significantly. 
When $q=0$, the residue in the integral~(\ref{defJ}) is 
$\tilde F_{\rm \Delta + p}(0)=1$, and so
\be \label{Jp0}
J(p,0)=-\frac{C_p}{C_0}\frac{\Gamma(2\Delta+2p)}{\Gamma^2(\Delta+p)}=-\frac{(2\Delta+2p-1)\Gamma(2\Delta+p-1)}{p! \Gamma^2(\Delta)}\ .
\ee 
When $p=0$ the sum~(\ref{a1}) reduces to a single term, which simplifies because some of the arguments 
of $\,_4F_3$ are equal:
\begin{equation} \label{J0p}
J(0,q)=-\frac{C_0 \Gamma(2\Delta)\Gamma^2(\Delta+q)}{C_q (q!)^2 \Gamma^4(\Delta)} F(-q,-q,2-2\Delta-2q,1)=-\frac{(2\Delta-1)\Gamma(2\Delta+q-1)}{q! \Gamma^2(\Delta)}
 \ .
\end{equation}

It may be worth mentioning that $J(p,q)$ can actually be expressed in terms of a single hypergeometric $\,_4F_3$
\begin{eqnarray}
  J(p,q) && = \frac{C_p}{C_q}\frac{\Gamma(2\Delta+2p)}{\Gamma(\Delta+p)^2}\frac{(-1)^{q}(1-q-\Delta)_{q}^2}{(2-2q-2\Delta)_{q}\Gamma(q+1)}
\nonumber\\ && \qquad\qquad \times
\,_4F_3(-q,1-p-\Delta,p+\Delta,2\Delta-1+q; 1,\Delta,\Delta; 1) \ .
\end{eqnarray}
However, this expression does not yield the special cases above in an obvious way.

It is also useful to determine the asymptotic expansion of $J(p,q)$ for $p\gg q,\Delta$.
The integral  $J(p,q)$ picks the term of order $z^q$ in
\begin{eqnarray}
&&(1-z)^p F(\Delta+p,\Delta+p,1,z) F(1-\Delta-q,1-\Delta-q,2-2\Delta-2q,z) \nonumber \\
&=&  \sum_{k_1,k_2,k_3=0}^\infty z^{k_1+k_2+k_3} \frac{(-p)_{k_1}}{k_1!}
\frac{(\Delta+p)_{k_2}^2}{(k_2!)^2}\frac{(1-\Delta-q)_{k_3}^2}{k_3! (2-2\Delta-2q)_{k_3}}\ .
\end{eqnarray}
This gives
\be
J(p,q)=-\frac{C_p \Gamma(2\Delta+2p)}{C_q \Gamma^2(\Delta+p)}
 \sum_{k_1,k_2,k_3=0}^\infty  \frac{(-p)_{k_1}(\Delta+p)_{k_2}^2(1-\Delta-q)_{k_3}^2}{k_1!(k_2!)^2 k_3! (2-2\Delta-2q)_{k_3}}
\delta_{k_1+k_2+k_3,q} \ ,
\ee 
where each term in the sum grows like $p^{k_1+2k_2}$ at large $p$. Therefore, the leading behavior of $J(p,q)$  comes from the term with $k_1=k_3=0$ and $k_2=q$. We obtain
\be
J(p,q)=-2 B_q p^{2\Delta+2q-1} + O(p^{2\Delta+2q-2})\ , \quad B_q = \frac{\Gamma(2\Delta+2q-1)}{ q! \Gamma^2(\Delta+q)\Gamma(2\Delta+q-1)}\ . \label{plargeq}
\ee
The regime $q\gg p,\Delta$ can easily be obtained using the relation (\ref{consist}),
\be
J(p,q) = -(2\Delta+2p-1)B_p q^{2\Delta+2p-2} \label{qlargeq} + O(q^{2\Delta+2p-3})\ .
\ee
\
We now give an outline of how to obtain the $L=0$ solution for the crossing equation in four dimensions. Setting $q=0$ in (\ref{crossingL0d4}) we find
\begin{equation}\label{crossingL0d4q0}
\gamma'(p,0)J^{(\Delta-1)}(p+1,0)-\gamma'(p-1,0)J^{(\Delta-1)}(p-1,0) = \gamma'(0,0)J^{(\Delta-1)}(1,p)
\end{equation}
$J^{(\Delta-1)}(p\pm 1,0)$ is obtained from (\ref{Jp0}), furthermore, using (\ref{consist})
\be
J^{(\Delta-1)}(1,p)  
=(\Delta-1)(2\Delta-1)(3-5\Delta+2((p-3)p+2p\Delta+\Delta^2))\frac{\Gamma(2\Delta+p-3)}{\Gamma(p+1)\Gamma^2(\Delta)}
\ee
We can then use a computer algebra program to iteratively solve (\ref{crossingL0d4q0}) for $\gamma(p,0)$ in terms of $\gamma(0,0)$. We conclude that the the general form of $\gamma(p,0)$ is given by (\ref{gammaL0}).

\sect{Nondegeneracy of $M(p)_{qr}$ and $M^{(\Delta-1)}(p)_{qr}$ }\label{nondegen}

In Sec.~4.2 and Sec.~7 we need to invert the $k \times k$ matrix
\begin{equation}
M(p)_{qr} = J(p+2r,q)\ ,\quad q,r = 0, \ldots, k-1
\end{equation}
at fixed $p$.
Suppose that for some constants $v_r$, $\sum_{r=0}^{k-1} M(p)_{qr} v_r = 0$ for all $0 \leq q\leq k-1$.  In other words,
\begin{eqnarray}
&&\oint_C \frac{dz}{2\pi i} \frac{\sigma(z)}{z^{q+1}}   \frac{F_{1 -\Delta - q}(z)}{(1-z)^{\Delta}} = 0\ , 
\quad 0 \leq q\leq k-1 \ ,
\nonumber\\
&&\sigma(z) = \sum_{r=0}^{k-1} v_r C_{\Delta + p + 2r} (1-z)^{\Delta+p+2r} \tilde F_{\Delta + p + 2r}(z)\ .
\end{eqnarray}
An iterative argument starting from $q=0$ shows that the contour integral vanishes iff the first $k$ terms in the Taylor series for $\sigma(z)$ at the origin vanish.

Define $\tilde{\cal D} = (1-z)^2 \partial_z z \partial_z$, which is related to $\cal D$ from Eq.~(\ref{eigen})  by $z \to 1-z$.  It follows from the hypergeometric equation that
\begin{equation}
\tilde{\cal D}(1-z)^{\Delta+p+2r} \tilde F_{\Delta + p + 2r}(z) = (\Delta + p + 2r)(\Delta + p + 2r - 1) (1-z)^{\Delta + p + 2r} \tilde F_{\Delta + p + 2r}(z)\ ,
\end{equation}
and so 
\begin{equation}
\left( \prod_{r=0}^{k-1} [ \tilde{\cal D} - (\Delta + p + 2r)(\Delta + p + 2r - 1) ] \right) \sigma(z) = 0\ . \label{wot}
\end{equation}
Now, the differential operator in this equation contains a term $(\partial_z z \partial_z)^k$, for which
\begin{equation}
(\partial_z z \partial_z)^k z^m = \frac{ \Gamma(m+1)^2}{\Gamma(m - k + 1)^2} z^{m-k}\ . 
\end{equation}
(Note that this vanishes for $m$ a nonnegative integer less than $k$.)  All other terms in the differential operator would give higher powers of $z$.  If $\sigma(z)$ has leading term $z^m$ with $m \geq k$, then it follows that there is a nonzero term of order $z^{m-k}$ on the left side of Eq.~(\ref{wot}), which is a contradiction.  Since we have already seen that there are no terms in $\sigma(z)$ with $m < k$, it follows that $\sigma(z)$ vanishes identically, and so do the $v_r$.  Thus, $M(p)_{qr}v_r = 0$ implies that $v_r=0$, QED.

In $d=4$ we need to invert a similarly defined matrix
\begin{equation}
M^{(\Delta-1)}(p)_{qr} = J^{(\Delta-1)}(p+2r+1,q)\ ,\quad q,r = 0, \ldots, k-1
\end{equation}
at fixed $p$. Suppose again that there exist constants $v_r$ such that $\sum_{r=0}^{k-1} M^{(\Delta-1)}(p)_{qr} v_r = 0$ for all $0 \leq q\leq k-1$. Then
\begin{eqnarray}
&&\oint_C \frac{dz}{2\pi i} \frac{\sigma(z)}{z^{q+1}}   \frac{F_{2-\Delta - q}(z)}{(1-z)^{\Delta}} = 0\ , 
\quad 0 \leq q\leq k-1 \ ,
\nonumber\\
&&\sigma(z) = \sum_{r=0}^{k-1} v_r C_{\Delta + p + 2r} (1-z)^{\Delta+p+2r} \tilde F_{\Delta + p + 2r}(z)\ .
\end{eqnarray}
This differs from $d=2$ only by the arguments in the second term of the integrand. This does not alter the previous line of reasoning and the proof holds as above; $M^{(\Delta-1)}(p)_{qr}v_r = 0$ implies that $v_r=0$.

\label{mqr}

\sect{Vanishing of $H(z)$}
\label{Hz}

After using the solution (\ref{L=0}), the expression (\ref{Hofz}) for $H(z)$ which appeared in the calculation of the $c(n,l)$ becomes
\be
H(z)=\gamma(0,0)\left[ F_{\Delta}(1-z)+\sum_{n=0}^\infty \frac{\partial}{\partial n}
 \left(C_n^2 \frac{(2\Delta-1)\Gamma(2\Delta+2n)}{(2\Delta+2n-1)\Gamma^2(\Delta+n)}
z^n F_{\Delta+n}(z)\right)\right]\ .
\ee
The infinite sum can be written as a countour integral
\ba
 \int \frac{dn}{2i\tan(\pi n)} \frac{\partial}{\partial n}
 \left(C_n^2 \frac{(2\Delta-1)\Gamma(2\Delta+2n)}{(2\Delta+2n-1)\Gamma^2(\Delta+n)}
z^n F_{\Delta+n}(z)\right)\ ,
 \ea
using the Sommerfeld-Watson transform.
Integrating by parts and changing variable, $n=(1-2\Delta+i\mu)/2$, we can rewrite the expression as an 
integral over real values of $\mu$ (assuming $\Delta>1/2$).
This gives
\ba
-\f{(2\Delta-1)}{4\pi\Gamma^4(\Delta)} \int d\mu\,  \Gamma^2(\Delta-\frac{1-i\mu}{2})\Gamma^2(\Delta-\frac{1+i\mu}{2})
 \frac{\Gamma^2(\frac{i\mu+1}{2})}
 { \Gamma(i\mu)}
 z^{\frac{i\mu+1}{2}-\Delta} F_{\frac{i\mu+1}{2}} (z) \ . 
\ea
We have checked numerically that this integral is precisely equal to $-F_{\Delta}(1-z)$ and therefore $H(z)$ vanishes.

\sect{Explicit Conformal Partial Wave Expansions in 2d}
\label{expCPW}
The conformal partial wave expansion (\ref{CPWexp}) of the correlator (\ref{AL2k4}) in $d=2$ can be found using a series expansion 
of the D-function. It reads
\ba
\gamma(n,0)&=& \frac{P_6(n)}{(2 n+2 \Delta -3) (2 n+2 \Delta -1) (2 n+2 \Delta
   +1)}\\
   \gamma (n,2)&=&\frac{(n+1) (n+2) (n+\Delta ) (n+\Delta +1) (n+2 \Delta
   -1) (n+2 \Delta )}{2 (2 n+2 \Delta -1) (2 n+2 \Delta +1) (2 n+2 \Delta +3)}
\ea
where
\ba
P_6(n)&=& 7 n^6+21 (2 \Delta -1) n^5+\left(99 \Delta ^2-93
   \Delta +16\right) n^4\\ &&+(2 \Delta  (\Delta  (58 \Delta -75)+20)+3) n^3+(\Delta  (\Delta  (38
   \Delta  (2 \Delta -3)+31)+11)-5) n^2 \nonumber \\ &&+2 \Delta ^3 (2 \Delta  (8 \Delta -13)+9) n+2 \Delta
   ^4 (4 (\Delta -2) \Delta +3)\nonumber 
\ea
and  $\gamma(n,l)=0$ for $l>2$. The $c(n,l)$ are given by equation (\ref{cdg}).

For the correlator (\ref{AL2k6}) the conformal partial wave expansion in $d=2$ is given by
\ba
\gamma(n,0)&=& \frac{P_8(n)}{ (2 n+2 \Delta -3) (2 n+2 \Delta -1) (2 n+2 \Delta +1)}\\
   \gamma (n,2)&=&\frac{(n+1) (n+2) (n+\Delta ) (n+\Delta +1) (n+2 \Delta -1) (n+2 \Delta )
  }{2 (2 n+2 \Delta -1) (2 n+2 \Delta
   +1) (2 n+2 \Delta +3)} 
      \\
   &   \times &  
   \left(3 n^2+(6\Delta +3) n+\Delta  (6 \Delta +7)+2\right)\nonumber 
\ea
where
\ba
P_8(n)&=&-3 n^8+12 (1-2 \Delta ) n^7+((100-57 \Delta ) \Delta -10) n^6\\&& -3 (\Delta  (\Delta  (2
   \Delta -89)+36)+4) n^5+(\Delta  (3 \Delta  (\Delta  (58 \Delta +93)-99)-14)+17) n^4\nonumber \\ &&
  +\Delta
    (\Delta  (4 (\Delta -1) \Delta  (72 \Delta +89)+13)+56) n^3  \nonumber \\ 
    &&+(\Delta  (\Delta  (\Delta  (2
   \Delta  (4 \Delta  (26 \Delta -7)-115)+25)+58)-10)-4) n^2 \nonumber \\
   &&+2 \Delta ^3 (2 \Delta  (\Delta 
   (4 \Delta  (5 \Delta -2)-23)+2)+9) n \nonumber 
   \\ &&+2 \Delta ^4 \left(8 \Delta ^4-4 \Delta ^3-14 \Delta
   ^2+\Delta +3\right)\nonumber 
\ea
and $\gamma(n,l)=0$ for $l>2$. The $c(n,l)$ are again given by equation (\ref{cdg}).

\end{appendix}

\end{document}